\documentclass[]{jfm}

\usepackage{svg}
\usepackage{url}
\usepackage{array}

\usepackage{enumitem}
\setlist[itemize,1]{leftmargin=*}
\setlist[itemize,2]{leftmargin=2em}

\usepackage{tabularx}

\usepackage{graphicx}
\usepackage{newtxtext}
\usepackage{newtxmath}
\usepackage{natbib}
\usepackage{hyperref}
\hypersetup{
    colorlinks = true,
    urlcolor   = blue,
    citecolor  = black,
}

\newcommand{\RomanNumeralCaps}[1]

\title{Scattering of Capillary-Gravity Waves by a Fixed, Semi-immersed Cylindrical Barrier with Contact Line Dissipation}

\author{Guoqin Liu 
 \and Likun Zhang 
   \corresp{\email{zhang@olemiss.edu}}}

\affiliation{National Center for Physical Acoustics and Department of Physics and Astronomy, University of Mississippi, University, MS 38677, USA}

\begin{document}
\maketitle

\begin{abstract}
The scattering of surface waves by structures intersecting liquid surfaces is fundamental in fluid mechanics, with prior studies exploring gravity, capillary, and capillary-gravity wave interactions. This paper develops a semi-analytical framework for capillary-gravity wave scattering by a fixed, horizontally placed, semi-immersed cylindrical barrier. Assuming linearized potential flow, the problem is formulated with differential equations, conformal mapping, and Fourier transforms, resulting in a compound integral equation framework solved numerically via the Nyström method. An effective-slip dynamic contact line model accounting for viscous dissipation links contact line velocity to deviations from equilibrium contact angles, with fixed and free contact lines of no dissipation as limiting cases. The framework computes transmission and reflection coefficients as functions of the Bond number, slip coefficient, and barrier radius, validating energy conservation and confirming a $90^\circ$ phase difference between transmission and reflection in specific limits. A closed-form solution for scattering by an infinitesimal barrier, derived using Fourier transforms, reveals spatial symmetry in the diffracted field, reduced transmission transitioning from gravity to capillary waves, and peak contact line dissipation when the slip coefficient matches the capillary wave phase speed. This dissipation, linked to impedance matching at the contact lines, persists across a range of barrier sizes. These results advance theoretical insights into surface-tension-dominated fluid mechanics, offering a robust theoretical framework for analyzing wave scattering and comparison with future experimental and numerical studies.

\end{abstract}


{\bf MSC Codes }  {\it(Optional)} Please enter your MSC Codes here

\section{Introduction}
\label{sec:Introduction}

\begin{figure}
    \centering
    \includegraphics[width=1\textwidth]{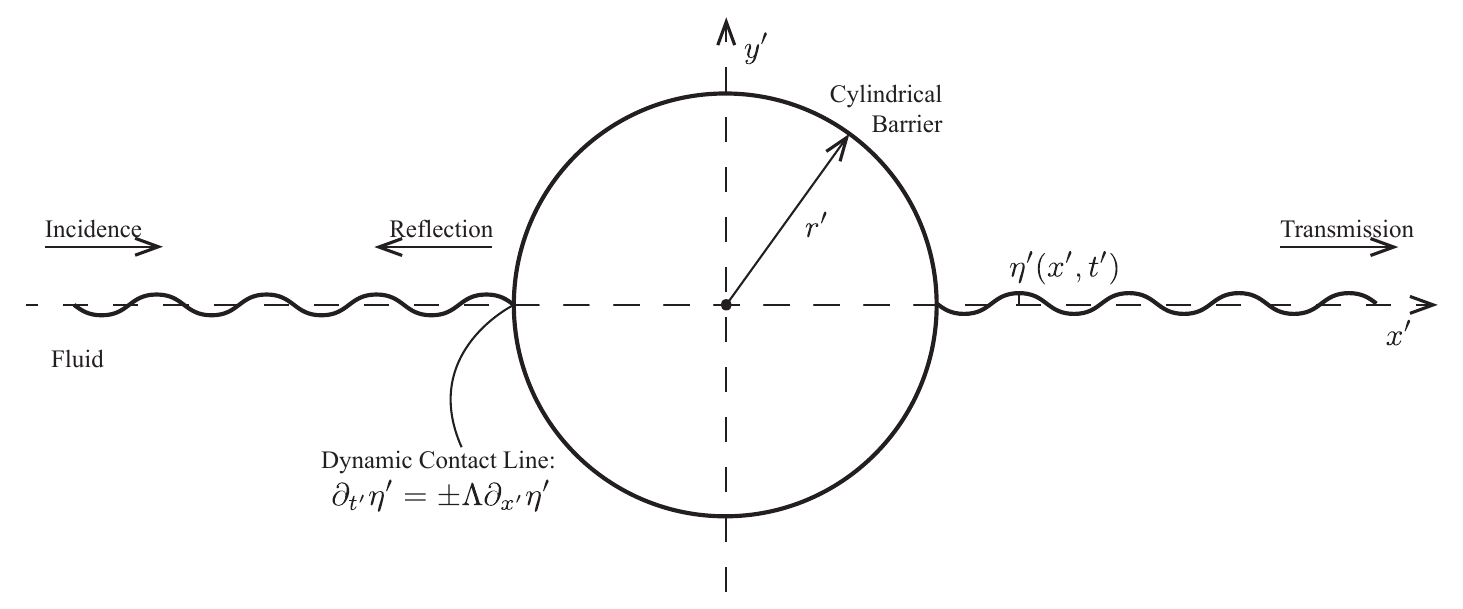}
    \caption{Schematic for capillary-gravity surface waves scattered by a fixed, semi-immersed, horizontal cylindrical barrier on a flat surface with a dynamic contact line.}
    \label{fig:schematic}
\end{figure}

The scattering of surface waves by barriers on liquid surfaces is a classical problem in fluid mechanics, with applications in coastal engineering, fluid containment, and wave energy systems. While scattering of gravity waves is typically observed in large-scale systems, scattering of capillary waves can occur in small-scale fluid systems. These systems often involve minimal surface-intersecting structures and have applications in laboratory setups and in the control and containment of capillary liquids, such as the stabilization of liquid globes by a wire loop \citep{ref:Pettit-2005}, helical-wire-stabilized liquid cylinders \citep{ref:Lowry-Thiessen-2007}, microstructured surfaces with pillars \citep{ref:Rothstein-2010}, and open capillary channels in biotechnology \citep{berthier2019open}. Capillary-gravity surface waves arise where both gravity and surface tension influence wave dynamics. While extensive studies have investigated gravity wave scattering, studies of capillary-gravity wave scattering remain limited, either experimental measurements \citep{ref:ParkLiuChan-2012,Guillaume2016,wang2023laboratory} or theoretical analysis \citep{packham1968capillary,Rhodes1982,RhodesR1996, ref:Hocking-1987,ref:Hocking-1990,zhang2013}. 

Theoretical analysis of capillary-gravity wave scattering is challenging due to the effect of surface tension, which raises the order of the surface boundary conditions \citep{ref:Hocking-model1-1987, ref:Henderson-Miles-1994} and requires an additional boundary condition at the contact lines—the three-phase junction where solid, liquid, and gas meet \citep{jiang2004contact, ref:Nicolas-2005, ref:Kidambi-2007, huang2020streaming, kim2020capillary}. Accounting for contact line motion is crucial for accurately modeling the scattering and other capillary phenomena in various fluid configurations, such as sessile drops \citep{ref:stick-slip-2004,vukasinovic2007dynamics, ref:Fayzrakhmanova-2009, sharp2012resonant}, bubbles \citep{harazi2019acoustics}, liquid bridges \citep{ref:Marr-Lyon2001, morse1996capillary}, and rivulets \citep{davis1980moving}. Prior experimental studies have shown that both fixed and mobile contact lines influence the dynamics, such as selectively exciting standing wave patterns in Faraday wave resonances \citep{zhang2024mechanically}, shifting the dispersion relation of traveling waves \citep{monsalve2022space}, and significantly altering surface wave scattering by modifying the contact line conditions \citep{ref:ParkLiuChan-2012,Guillaume2016,wang2023laboratory}. The few previous theoretical analyses of capillary-gravity wave scattering with contact line conditions focused primarily on scattering from vertical cliffs \citep{Rhodes1982}, infinitesimally thin barriers \citep{RhodesR1996, ref:Hocking-1987,zhang2013}, vertical cylinders \citep{ref:Hocking-1990}, and porous plates  \citep{gorgui1993capillary,rhodes1997waves,chakrabarti1998effect}. In this paper, we develop a semi-analytical solution for the scattering of capillary-gravity waves by a fixed, semi-immersed cylindrical barrier, applying a linear dynamic contact line condition at the three-phase junction (Fig.~\ref{fig:schematic}).

The scattering of gravity waves by barriers has been a cornerstone of hydrodynamic research, with foundational studies by \citet{Dean1945} and \citet{Ursell1947} providing critical insights into wave interactions with submerged and surface-piercing structures. Dean analyzed the reflection of surface waves by a submerged plane barrier, while Ursell focused on the effect of a fixed vertical barrier on wave transmission and reflection in deep water, laying the groundwork for subsequent advancements. 
Built upon these seminal works, several studies were conducted between 1950 and 2000 to deepen the understanding of gravity wave scattering \citep{mei1989applied}. \citet{wehausen1960handbuch} provided a comprehensive review of surface waves, synthesizing earlier findings, including those of Dean and Ursell. Classic work such as \citet{newman1965propagation} analyzed the propagation of gravity waves past a step bottom, providing foundational insights that continue to inform modern studies of wave scattering. Further studies, such as \citet{liu1982wave}, employed the boundary integral equation method to investigate the scattering characteristics of gravity waves by vertical and inclined thin breakwaters, focusing on numerical solutions for reflection and transmission coefficients. Similarly, \citet{sahoo2000trapping} examined wave trapping and generation by submerged vertical porous barriers, employing eigenfunction expansion and least-squares methods to analyze reflection coefficients, dynamic pressure distributions, and wave amplitudes for various porous-effect parameters.

Recent research has extended these insights to more complex barrier designs, including porous, submerged, and resonator-based configurations, with applications in wave energy dissipation and coastal protection. For instance, \citet{chanda2022scattering} investigated flexural-gravity wave interaction with submerged porous barriers under an ice-sheet modeled as a thin elastic plate, highlighting the influence of structural parameters on energy loss and wave forces. \citet{chanda2024flexural} explored flexural-gravity wave scattering by porous barriers in various configurations, introducing a scattering matrix to analyze mode conversion, Bragg resonance, and energy dissipation. \citet{manam2014membrane} used integral equations to compute reflection coefficients for membrane-coupled wave scattering by barriers with gaps, demonstrating the role of structural parameters in wave reflection. \citet{lorenzo2023attenuating} reported on a metamaterial-inspired wave attenuation device using submerged cylindrical pendula, showcasing resonance effects and Bragg scattering as mechanisms for significant energy dissipation and reflection. This evolving body of knowledge continues to inform the design of efficient and environmentally sustainable wave mitigation structures.

The study of surface wave scattering by cylindrical structures has been another cornerstone of hydrodynamic research, with applications in offshore engineering and wave energy management. A foundational work by \citet{martin1983scattering} introduced a rigorous analytical framework to investigate gravity wave scattering by a semi-immersed fixed circular cylinder, highlighting the significance of diffraction effects and providing insights into reflection, transmission, and hydrodynamic forces. Numerous studies have explored more complex systems, including floating cylinders \citep{porter2009water}, porous media \citep{zhuang2023nonlinearity}, and metamaterial-based wave manipulation \citep{zheng2024wave,zhu2024controlling}. Recent advancements in computational methods, including high-fidelity numerical models, have provided insights into nonlinear wave-structure interactions, such as wave attenuation by floating breakwaters \citep{peng2023computational}. Complementary numerical investigations have further explored novel structural configurations for enhanced wave energy dissipation, analyzing performance over a broad range of parameters \citep{panduranga2021water}.

The introduction of surface tension into wave scattering problems has been examined under diverse conditions including capillary-gravity wave scattering by surface tension gradient \citep{gou1993capillary,chou1995capillary}, by a non-uniform current \citep{trulsen1993double},  by surface convection zones \citep{kistovich2005reflection}, by a completely submerged porous elastic plate \citep{singh2022scattering}, and by bottom modulations \citep{mohapatra2016effect,meylan2024scattering}. In these contexts, no structure intersects the free surface, so contact line effects do not arise. Scattering of capillary-gravity waves by surface-intersecting barriers with contact line effects was foundationally explored by \citet{ref:Evans-1968}, \citet{packham1968capillary}, \citet{Rhodes1982, RhodesR1996}, \citet{ref:Hocking-1987}, and more recently, \citet{zhang2013}. Evans and Packham examined the impact of surface tension on wave reflection by a vertical barrier, establishing critical edge conditions that have influenced subsequent studies. Rhodes-Robinson and Hocking independently investigated wave reflection using a linear dynamic contact line model following different approaches. \citet{zhang2013} introduced dissipation models for capillary waves scattered through thin barriers, advancing the field’s understanding of capillary wave energy dissipation mechanisms at dynamic contact lines. Another related work by \citet{ref:Hocking-1990} analytically examined capillary-gravity wave scattering by a vertical cylinder using the same linear dynamic contact line model adopted here.

There are a few experimental investigations relevant to capillary-gravity wave scattering from barriers, particularly regarding the role of dynamic contact lines. \citet{ref:ParkLiuChan-2012} investigated solitary wave reflection off vertical walls, using particle tracking velocimetry (PTV) to analyze the velocity field near the moving contact line. Their findings revealed complex flow phenomena, including surface rolling, boundary layer flow reversal, and the formation of jets and eddies near the meniscus. \citet{Guillaume2016} investigated the reflection of capillary-gravity surface waves by a plane vertical barrier, demonstrating that meniscus size significantly affects wave reflection—with a pinned contact line yielding nearly double the reflected energy compared to a fully developed meniscus—and introduced an acousto-optic-inspired acoustic measurement system to accurately measure wave amplitude, frequency, and propagation direction. \citet{wang2023laboratory} performed experiments using acoustic measurements to quantify the effects of barrier depth, width, and wave frequency on scattering, validating theoretical predictions of contact line dynamics in realistic fluid configurations and revealing the significant meniscus effect in capillary-gravity wave scattering.

Classical no-slip conditions lead to singular stresses, necessitating modified boundary conditions that better represent microscopic physics. Foundational works by \citet{dussan1979spreading} and others \citep{cox1986dynamics1, cox1986dynamics2, voinov1976hydrodynamics, tanner1979spreading, de1985wetting} clarified that dynamic contact lines cannot be handled by classical slip-free models. Dynamics of contact lines can be modeled by the interplay between the apparent contact angle and contact-line velocity in the macroscopic regime, governing behaviors independent of microscopic complexities. We adopt a linearized dynamic contact line condition by \citet{ref:Hocking-1987} that bridges microscopic considerations with macroscopic wave behavior by introducing an “effective-slip” model. Hocking’s effective-slip model links the contact line velocity to the deviation of the contact angle from its equilibrium value, scaled by a slip coefficient with units of velocity. This slip coefficient encapsulates the interfacial properties of the solid–liquid boundary. At limiting extremes, a zero slip coefficient enforces ``fixed'' contact lines, while an infinite slip coefficient ensures ``free'' contact lines that maintain the contact angle with no energy dissipation. Between these extremes, finite slip coefficients produce boundary work as the contact line moves, converting a portion of wave energy into dissipation \citep{ref:Hocking-model1-1987, ref:Hocking-1987}. The linear dynamic contact line model has been extensively used in analyses of various surface tension phenomena, such as behavior of surface-wave damping in closed basins \citep{ref:Miles-1967,ref:Kidambi-2009}, oscillating droplets on solid plates \citep{ref:Lyubimov-2006}, and forced oscillations of a cylindrical droplet bounded in the axial direction by rigid planes \citep{alabuzhev2016axisymmetric}.

Our work aims to establish a milestone theoretical framework for surface wave scattering in fluid mechanics, by incorporating surface tension and a linear dynamic contact line condition into scattering of the waves by a cylindrical barrier with a nonzero cross-sectional dimension. While \citet{ref:Hocking-1987} included surface tension in the scattering but assumed an infinitesimal thin barrier, and \citet{martin1983scattering} considered cylindrical barrier geometry but for gravity waves, the integration of both surface tension and cylindrical barrier geometry in our study has constituted a challenging problem in fluid mechanics. In the scattering problem considered by \citet{ref:Hocking-1990}, a vertical cylinder geometry allows the wave field to be decomposed into an incident field plus a localized scattered field that vanishes at infinity. In our problem, however, a horizontally placed cylindrical barrier extends to infinity, rendering the scattered wave field (both transmitted and reflected components) unbounded, thus posing a more demanding analytical challenge. 

Under the assumption of a linearized problem for an inviscid, incompressible, and irrotational fluid, we employ techniques such as differential equations, conformal mapping, and Fourier transforms to solve the coupled boundary value problems and formulate the entire problem as a compound integral equation problem. The resulting integral equations are simplified using integral equation techniques and numerically discretized via the Nyström method, enabling the computation of transmission and reflection coefficients of the scattered waves. 

The paper is organized as follows. The linearized problem is formulated and divided into two boundary value problems (BVP) in \S~\ref{sec:Formulation}. The two BVPs are solved and converted to coupled integral equations using conformal mapping and Fourier transform in \S~\ref{sec:Solving Boundary Value Problems}. The integral equations are combined and solved using even-odd separation in \S~\ref{sec:Solving Integral Equations}. The validation of the model by comparing results with previous work in gravity waves is presented in \S~\ref{sec:Gravity Waves}. The wave phase and spatial symmetry when no dissipation for capillary-gravity waves is addressed in \S~\ref{sec:No Dissipation}. The capillary-gravity wave scattering by an infinitesimal barrier is solved analytically using Fourier cosine and sine transform (FCT, FST) in \S~\ref{sec:Infinitesimal Barrier}. The calculated results for fixed and free contact lines are presented in \S~\ref{sec:Capillary-Gravity Waves without Dissipation} for barriers ranging from infinitesimal to having a finite cross-sectional dimension. The contact-line dissipation is being addressed in \S~\ref{sec:dissipation}. All results are summarized and discussed in \S~\ref{sec:Summary}.

\section{Formulation: Linearized Problem \& Contact Line Model}
\label{sec:Formulation}

The analysis begins by considering an equilibrium configuration where a horizontally placed cylindrical barrier of radius $r'$ is fixed and semi-immersed in an infinitely deep liquid (see Fig.~\ref{fig:schematic}). The infinitely deep liquid refers to the case when the liquid depth is much larger than the wavelength. A Cartesian coordinate system $(x', y')$ is defined, with the origin located at the center of the cylindrical barrier. The cylinder is uniformly extended along the third dimension $z'$, corresponding to the scenario that the cylinder length is much larger than the wavelength. The liquid, characterized by density $\rho$, surface tension $\sigma$, and unperturbed pressure $p_0' = P_0' - \rho g y'$ ($P_0'$ is the atmospheric pressure), forms a static contact angle of $90^{\circ}$ with the barrier. This setup is used to investigate the scattering of time-harmonic capillary-gravity waves—defined by angular frequency $\omega$ and wavenumber $k$ —incident from $x' \rightarrow -\infty$.

For an inviscid, incompressible, and irrotational fluid, the fluid motion is governed by the continuity equation, the Navier-Stokes equation, alongside the kinematic and dynamic conditions at the free surface:
\begin{subequations}\label{eq:original}
\begin{align}
\bnabla' \bcdot \boldsymbol{v}' &= 0, \quad &&\forall (x', y') \in \Omega', \\
\partial_{t'} \boldsymbol{v}' + (\boldsymbol{v}' \bcdot \bnabla') \boldsymbol{v}' &= - \rho^{-1} \bnabla' p', \quad &&\forall (x', y') \in \Omega', \\
\partial_{t'} \eta' + \boldsymbol{v}' \bcdot \bnabla' \eta' &= \boldsymbol{\hat{n}}' \bcdot \boldsymbol{v}', \quad && \text{at } |x'| > r', y' = \eta', \\
p_0' + p' &= P_0' + \sigma \bnabla' \bcdot \boldsymbol{\hat{n}}', \quad && \text{at } |x'| > r', y' = \eta',
\end{align}
\end{subequations}
where $\boldsymbol{v}' = \boldsymbol{v}'(x', y', t')$ is the velocity field, $p' = p'(x', y', t')$ is the perturbed pressure of the wave field, $\eta' = \eta'(x', t')$ is the surface elevation with respect to its equilibrium position, and $\Omega' = \{(x', y') \in \mathbb{R}^2 \mid y' < 0 \text{ and } x'^2 + y'^2 > r'^2\}$ defines the liquid domain. $\boldsymbol{\hat{n}}' = \bnabla'(y'-\eta')/|\bnabla'(y'-\eta')|$ is the normal vector of the free surface pointing outward of the liquid domain, and $\bnabla' \bcdot \boldsymbol{\hat{n}}'$ represents the curvature of the free surface. 

Introducing the velocity potential with $\boldsymbol{v}' = \bnabla' \phi'$, and considering a linearized problem, \eqref{eq:original} simplifies to:
\begin{subequations}\label{eq:original_2}
\begin{align}
\nabla'^2 \phi' &= 0, \quad &&\forall (x', y') \in \Omega', \label{eq:laplace} \\
\partial_{t'} \phi' &= - p' / \rho, \quad &&\forall (x', y') \in \Omega', \label{eq:dynamic_phi} \\
\partial_{y'} \phi' &= \partial_{t'} \eta' , \quad &&\text{at } |x'| > r', y' = 0, \label{eq:kinematic} \\
p' &= - \sigma \partial_{x'}^2 \eta' + \rho g \eta' , \quad &&\text{at } |x'| > r', y' = 0. \label{eq:dynamic_eta}
\end{align}
\end{subequations}
To ensure the validity of the linearization, we assume that the perturbation from the equilibrium configuration is small. Specifically, the surface elevation $\eta'$ must satisfy $k \eta' \ll 1$.

Besides, the problem has to be solved subject to the non-penetration condition at the barrier:
\begin{equation}
\boldsymbol{\hat{n}}' \bcdot \bnabla' \phi' = 0, \qquad \text{at } |x'| < r', y' = - \sqrt{r'^2 - x'^2}, \label{eq:barrierBC}
\end{equation}
where $\boldsymbol{\hat{n}}'$ represents the normal vector at the cylindrical barrier surface.

The edge conditions imposed on $\eta'$ at the three-phase junction are given by the linear dynamic contact line condition from \citet{ref:Hocking-1987}:
\begin{equation}
\partial_{t'} \eta' = \pm \Lambda' \partial_{x'} \eta', \qquad \text{at } x' = \pm r', \label{eq:dcl}
\end{equation}
where the slip coefficient $\Lambda'$ has dimensions of velocity and is taken to be a non-negative real-valued constant. The plus and minus signs correspond to the right and left sides of the barrier, respectively. 

The imposed edge condition given in \eqref{eq:dcl} governs the contact line’s behavior via the slip coefficient $\Lambda'$. The condition includes two special cases as its extremes: When $\Lambda' = 0$, the contact line is effectively fixed (i.e. $\partial_{t'}\eta' = 0$), termed as “fixed” contact lines; conversely, as $\Lambda' \to \infty$, the contact line becomes completely free to slide while maintaining a constant contact angle which is $90^\circ$ here (i.e. $\partial_{x'} \eta' = 0$), termed as “free” contact lines. There is no energy dissipation in these two extremes. Intermediate values of $\Lambda'$ between $0$ and $\infty$ represent dynamic contact lines involving energy dissipation.

Additionally, the free surface elevation at the contact line is assumed to be small compared to the barrier radius, specifically, $\eta' \ll r'$, ensuring the motion of the contact line along the $x$-axis remains negligible, and preserving the validity of the linear dynamic contact line model.

At the free surface of the liquid of infinite depth, the incident surface wave of angular frequency $\omega$ and wavenumber $k$ has the form of 
\begin{subequations}\label{eq:incident}
\begin{align}
\phi'_I ( x', y', t' ) &= \phi_A \mathrm{e}^{ k y' } \mathrm{e}^{ \mathrm{i}( k x' - \omega t' ) } , \\
\eta'_I ( x', t' ) &= \eta_A \mathrm{e}^{ \mathrm{i}( k x' - \omega t' ) } , 
\end{align}
where amplitude $\phi_A$ and $\eta_A$ has the linear relation
\begin{equation}
\phi_A = - \mathrm{i}\frac{\omega}{k} \eta_A,
\end{equation}
satisfying the kinematic boundary condition \eqref{eq:kinematic}, and $\omega$ and $k$ have the dispersion relation 
\begin{equation}
\omega^2 = \frac{ \sigma }{ \rho } k^3 + g k = \frac{ \sigma }{ \rho } k^3 (1 + B),
\end{equation}
\end{subequations}
satisfying the dynamic boundary condition \eqref{eq:dynamic_phi} and \eqref{eq:dynamic_eta}, where 
\begin{equation}\label{eq:B}
B = \frac{\rho g}{\sigma k^2},
\end{equation}
is the Bond number, characterizing the relative effect between gravity and surface tension. 

At large distances from the barrier, the surface elevation takes the far-field form:
\begin{subequations}\label{eq:farfield}
\begin{align}
\eta' &= \eta'_I + \eta'_R, \quad &\text{at } x' \rightarrow -\infty, \\
\eta' &= \eta'_T, \quad &\text{at } x' \rightarrow +\infty,
\end{align}
where 
\begin{align}
\eta'_R &= R \eta'_I (-x', t') = R \eta_A \mathrm{e}^{ \mathrm{i}( - k x' - \omega t' ) } ,\\
\eta'_T &= T \eta'_I (+x', t') = T \eta_A \mathrm{e}^{ \mathrm{i}( k x' - \omega t' ) } .
\end{align}
\end{subequations}
Here, $T$ and $R$ are the complex transmission and reflection coefficients, respectively, which will be determined for their dependence on slip coefficient $\Lambda'$, Bond number $B$, wavenumber $k$, and barrier radius $r'$.

For the linearized problem, where the incident wave \eqref{eq:incident} has the form of time harmonic oscillations $\mathrm{e}^{-\mathrm{i}\omega t'}$ and all governing equations including the dynamic contact line condition \eqref{eq:dcl} are homogeneous, the wave fields $\phi'$ and $\eta'$ naturally adopt the same form of time harmonic oscillations:
\begin{subequations}
\begin{align}
\phi' ( x', y', t' ) &= \phi'(x', y') \mathrm{e}^{ - \mathrm{i}\omega t' } , \\
\eta' ( x', t' ) &= \eta'(x') \mathrm{e}^{ - \mathrm{i}\omega t' } .
\end{align}
\end{subequations}
The original spatiotemporal problem reduces to a purely spatial one for $\phi'(x', y')$ and $\eta'(x')$. 

The problem is transformed into a dimensionless form by introducing a new set of variables devoid of units: 
\begin{equation}
    t = \omega t',\quad (x, y, r) = k (x', y', r'),\quad \eta = \eta' / \eta_A,\quad \phi = \phi' / \phi_A,\quad p = p' / (\sigma k^2).
\end{equation}
This rescaling simplifies the mathematical expressions and highlights the fundamental dynamics independent of specific scales of time and space. Furthermore, we redefine the slip coefficient $\Lambda'$ in dimensionless terms as
\begin{equation}
\Lambda = \frac{\Lambda'}{c_p} = \frac{1}{\sqrt{1+B}} \sqrt\frac{\rho}{\sigma k} \Lambda',
\end{equation}
where $c_p = \omega/k = \sqrt{(1+B) \sigma k / \rho}$ is the phase speed of the capillary-gravity wave.

The smallness conditions $k \eta_A \ll 1$ and $k \eta_A \ll r$ remain central to the dimensionless formulation. They ensure that the problem is governed by linearized dynamics, simplifying the analysis while maintaining accuracy for small perturbations.

The problem then reduces to solving two coupled boundary value problems (BVP) for the velocity potential $\phi(x, y)$ and the surface elevation $\eta(x)$. For $\phi$, following \eqref{eq:laplace}, \eqref{eq:barrierBC}, and \eqref{eq:kinematic}, we solve:
\begin{equation}
\begin{cases}\label{eq:phiequ}
\displaystyle \nabla^2 \phi = 0, & \forall (x, y) \in \Omega, \\
\displaystyle \boldsymbol{\hat{n} \cdot \nabla} \phi = 0, & \text{at } |x| < r, y = - \sqrt{r^2 - x^2}, \\
\displaystyle \boldsymbol{\hat{n} \cdot \nabla} \phi = \eta, & \text{at } |x| > r, y = 0,
\end{cases}
\end{equation}
where the domain $\Omega = \{(x, y) \in \mathbb{R}^2 \mid y < 0 \text{ and } x^2 + y^2 > r^2\}$ and $\boldsymbol{\hat{n}}$ represents the outward normal vector of the domain boundary. For $\eta$, following \eqref{eq:dynamic_phi}, \eqref{eq:dynamic_eta}, \eqref{eq:dcl} and \eqref{eq:farfield}, we solve, on the surface of the incident side ($x < -r$):
\begin{subequations}\label{eq:etaequ}
\begin{equation}
\begin{cases}\label{eq:etaequ1}
\displaystyle \frac{\mathrm{d}^2 \eta}{\mathrm{d}x^2} - B \eta = - (1+B) \phi \big|_{y=0}, & \forall x < -r,  \\
\displaystyle \frac{\mathrm{d} \eta}{\mathrm{d}x} = \mathrm{i}\Lambda^{-1} \eta, & \text{at } x = -r, \\
\displaystyle \eta \rightarrow \mathrm{e}^{\mathrm{i}x} + R \mathrm{e}^{-\mathrm{i}x}, & \text{as } x \rightarrow -\infty,
\end{cases}
\end{equation}
and on the surface of the transmitted side ($x > r$):
\begin{equation}
\begin{cases}\label{eq:etaequ2}
\displaystyle \frac{\mathrm{d}^2 \eta}{\mathrm{d}x^2} - B \eta = - (1+B) \phi \big|_{y=0}, & \forall x > r,  \\
\displaystyle \frac{\mathrm{d} \eta}{\mathrm{d}x} = - \mathrm{i}\Lambda^{-1} \eta, & \text{at } x = r, \\
\displaystyle \eta \rightarrow T \mathrm{e}^{ \mathrm{i}x }, & \text{as } x \rightarrow +\infty.  
\end{cases}
\end{equation}
\end{subequations}

Solving these BVPs allows us to determine the complex transmission ($T$) and reflection ($R$) coefficients. These coefficients are dependent on three key dimensionless parameters: the Bond number $B$, the dimensionless radius of the barrier $r$, and the dimensionless slip coefficient $\Lambda$. This formulation highlights the interplay between fluid dynamics and the physical geometry of the barrier.

\section{Solving Boundary Value Problems}
\label{sec:Solving Boundary Value Problems}

\subsection{Expressing $\eta$ in terms of $\phi$}

To solve the BVPs for $\eta$, as specified in \eqref{eq:etaequ1} and \eqref{eq:etaequ2}, we express $\eta$ in terms of the boundary condition $\phi|_{y=0} = \phi(x, 0)$. These BVPs are modeled as inhomogeneous second-order linear differential equations (ISOLDEs), which are solved using the method of variation of constants \citep{hassani2013mathematical}. With detailed derivations provided in Appendix~\ref{app:solve_eta}, the solution for $\eta$ is given by:
\begin{equation}
\eta = 
\begin{cases}\label{eq:etasol}
\displaystyle \int^{- r}_{- \infty} K(x, x') \phi(x', 0) \: \mathrm{d} x', & \forall x < -r,  \\
\displaystyle \int^{+ \infty}_{r} K(x, x') \phi(x', 0) \: \mathrm{d} x', & \forall x > r,
\end{cases}
\end{equation}
where the dynamic kernel $K(x, x')$ is defined as:
\begin{equation}
K(x, x') = \frac{1+B}{2\sqrt{B}} \left( -\mathrm{e}^{-2\mathrm{i}\zeta} \mathrm{e}^{\sqrt{B}(2r - |x + x'|)} + \mathrm{e}^{-\sqrt{B}|x - x'|} \right), \label{eq:K}
\end{equation}
in which the parameter $\zeta$ is defined as:
\begin{equation}
\zeta = \tan^{-1} \left( \sqrt{B}\Lambda \right) = \tan^{-1} \left( \sqrt{\frac{B}{1+B}} \sqrt{\frac{\rho}{\sigma k}} \Lambda' \right). \label{eq:theta}
\end{equation}
This formulation captures the spatial dependencies within the problem, allowing $\eta$ to be directly related to the boundary values of $\phi$.

The parameter $\zeta$ in \eqref{eq:theta} is introduced for mathematical convenience, simplifying the exponential expressions in the dynamic kernel $K$ in \eqref{eq:K}. This parameter $\zeta$ eventually encodes the dependence of the problem on the slip coefficient $\Lambda'$. It ranges from $0$ to $\pi/2$ as the slip coefficient $\Lambda'$ varies from $0$ to $\infty$, unless the Bond number $B = 0$. From now on, $\zeta$ replaces $\Lambda$ to act as the dimensionless slip parameter in the dynamic contact line conditions. It is worth pointing out that, with the inverse trigonometric function arctangent in \eqref{eq:theta}, this slip parameter $\zeta$ may be interpreted as an ``angle.'' However, it should not be confused with the contact angle at the edges, whose deviation from the static contact angle is given by $\tan^{-1} (\partial_{x'}\eta')$.

\subsection{Expressing $\phi$ in terms of $\eta$}

\subsubsection{Conformal Mapping}
\label{ssec:Conformal Mapping}

\begin{figure}
    \centering
    \includegraphics[width=\textwidth]{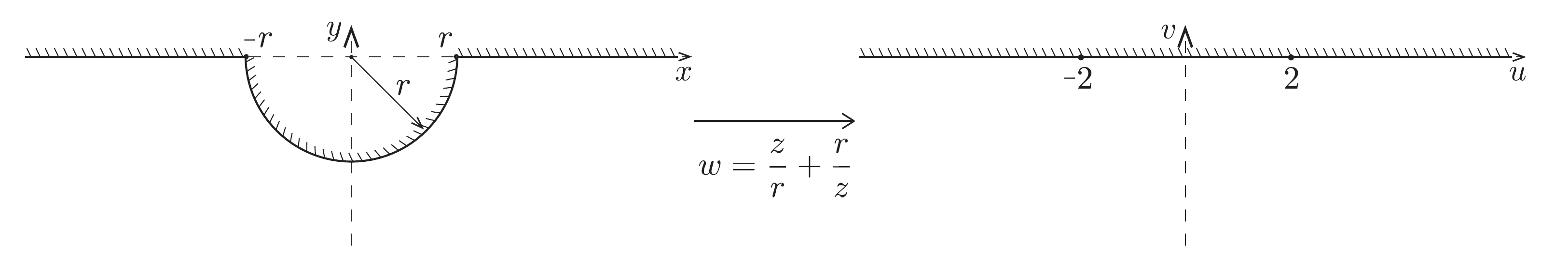}
    \caption{Schematic for the conformal mapping.}
    \label{fig:schematic_conformal}
\end{figure}

In addressing the BVP for $\phi(x, y)$ as defined in \eqref{eq:phiequ}, we employ a conformal mapping technique. This method transforms the liquid domain $\Omega$ into the lower half-space $H = \{(u, v) \in \mathbb{R}^2 \mid v < 0\}$. Specifically, the free surface and the lower half of the cylindrical barrier surface are mapped onto the real axis, as depicted in Fig.~\ref{fig:schematic_conformal}. Defining $z = x + \mathrm{i} y$ and $w = u + \mathrm{i} v$, the mapping from the $z$-plane to the $w$-plane is prescribed by:
\begin{equation}
w = \frac{z}{r} + \frac{r}{z},
\end{equation}
or, in component form, 
\begin{equation}\label{eq:conformal_components}
u = \frac{(x^2 + y^2 + r^2) x}{(x^2 + y^2) r}, \quad v = \frac{(x^2 + y^2 - r^2) y}{(x^2 + y^2) r}.
\end{equation}
This conformal mapping, known as the Joukowsky transform and historically used for applications like airfoil design \citep{joukowsky1910konturen}, is extended here to analyze capillary-gravity wave scattering.

In this transformed coordinate system, the free surface is defined by $v = 0, |u| > 2$, and the barrier surface by $v = 0, |u| < 2$. The transformation preserves the Laplace equation; thus the transformed BVP for $\psi$, related to $\phi$ by $\psi(u, v) = \phi(x, y)$, is expressed as:
\begin{equation}
\begin{cases}\label{eq:psiequ}
\displaystyle \nabla^2 \psi = 0, & \forall (u, v) \in H, \\
\displaystyle \boldsymbol{\hat{n} \cdot \nabla} \psi = g(u), & \text{at } v = 0.
\end{cases}
\end{equation}
The boundary function $g(u)$ is defined by:
\begin{equation}\label{eq:g2eta}
g(u) = 
\begin{cases}
\displaystyle 0, & \forall |u| < 2,  \\
\displaystyle \eta / \partial_x u(x, 0), & \forall |u| > 2.
\end{cases}
\end{equation}

\subsubsection{Fourier Transform}
\label{ssec:Fourier Transform}

Under normal circumstances, the boundary value problem (BVP) for $\psi(u, v)$ as outlined in \eqref{eq:psiequ}, being a Laplace equation with Neumann boundary conditions in the lower half-space, should be solvable using the Green's function approach. However, our situation presents a unique challenge due to the behavior of the field at infinity ($u \to \pm \infty$), where it does not vanish. This non-vanishing at infinity is mirrored by the corresponding Green's function for the Neumann boundary condition in the lower half-space, which similarly does not vanish as $u \to \pm \infty$. Consequently, the integral involving the Green's function multiplied by $\nabla \psi$ at infinity does not converge, rendering this approach infeasible.

We opted for a different approach, utilizing the Fourier transform method. This elegant technique circumvents the difficulties associated with the traditional Green's function method and provides a robust solution to the problem. 

We apply the following Fourier transform to \eqref{eq:psiequ},
\begin{equation}
    \hat{f}(\lambda) = \int_{-\infty}^{+\infty} f(u) \mathrm{e}^{-\mathrm{i}\lambda u} \: \mathrm{d} u,
\end{equation}
we obtain
\begin{equation}
\begin{cases}\label{eq:psiequfourier}
\displaystyle \frac{\partial^2}{\partial v^2} \hat{\psi}(\lambda, v) - \lambda^2 \hat{\psi}(\lambda, v) = 0, & \forall v < 0, \\
\displaystyle \frac{\partial}{\partial v} \hat{\psi}(\lambda, v) = \hat{g}(\lambda), & \text{at } v = 0,
\end{cases}
\end{equation}
where $\hat{\psi}$ and $\hat{g}$ are the Fourier transform of $\psi$ and $g$. This simplifies the problem to a homogeneous ODE with respect to $v$, allowing $\hat{\psi}(\lambda, v)$ to be expressed as:
\begin{equation}
\hat{\psi}(\lambda, v) = \frac{\hat{g}(\lambda)}{|\lambda|} \mathrm{e}^{|\lambda| v}, \quad \forall v < 0.
\end{equation}
An inverse Fourier transform yields the solution to \eqref{eq:psiequ}:
\begin{align}\label{eq:phi0sol_temp}
\psi(u, v) 
&= \frac{1}{2\upi} \int_{-\infty}^{+\infty} g(u') \left( \int_{-\infty}^{+\infty} \frac{1}{|\lambda|} \mathrm{e}^{|\lambda| v} \mathrm{e}^{\mathrm{i}\lambda (u - u')} d \lambda \right) \mathrm{d} u' .
\end{align}
By evaluating at $v = 0$, with detailed derivations provided in Appendix~\ref{app:solve_phi}, we derive the expression for $\phi(x,0) = \psi(u,0)$, giving:
\begin{align}\label{eq:phi0sol}
\phi(x, 0)
&= \left( \int_{-\infty}^{-r} + \int_{r}^{+\infty} \right) \, S (x, x') \eta(x') \: \mathrm{d} x' ,  \quad \forall |x| > r,
\end{align}
where the geometric kernel $S(x, x')$ is defined as:
\begin{equation}
S (x, x') = - \upi^{-1} \left( \ln{ \left| \frac{x}{r} + \frac{r}{x} - \frac{x'}{r} - \frac{r}{x'} \right| } + \gamma \right), \label{eq:S}
\end{equation}
where $\gamma$ is the Euler-Mascheroni constant ($= 0.57722 \cdots$). 

We have now solved the BVPs referred to in \eqref{eq:phiequ}-\eqref{eq:etaequ2}. Specifically, the solutions for $\eta$ in terms of $\phi(x, 0)$ are presented in \eqref{eq:etasol}, while the solution for $\phi(x, 0)$ in terms of $\eta$ is provided in \eqref{eq:phi0sol}. By combining \eqref{eq:etasol} and \eqref{eq:phi0sol}, we can derive a single integral equation for $\phi(x, 0)$. Solving this integral equation will allow us to determine the complex transmission ($T$) and reflection ($R$) coefficients.

\section{Solving Integral Equations}
\label{sec:Solving Integral Equations}

Let us denote $f(x) = \phi(x, 0)$ for simplicity. By combining \eqref{eq:etasol} and \eqref{eq:phi0sol}, we arrive at the following integral equation for $f(x)$, which must also satisfy specific far-field forms as $x \to \pm \infty$:
\begin{equation}
\begin{cases}\label{eq:fequ}
\displaystyle f(x) = \left( \int_{-\infty}^{-r} \int_{-\infty}^{-r} + \int_{r}^{+\infty} \int_{r}^{+\infty} \right) \, S (x, x') K(x', x'') f(x'') \: \mathrm{d} x' \: \mathrm{d} x'' ,  \quad \forall |x| > r,  \\
\displaystyle f(x) \rightarrow \mathrm{e}^{\mathrm{i}x} + R \mathrm{e}^{-\mathrm{i}x}, \qquad\ \text{as } x \rightarrow -\infty, \\
\displaystyle f(x) \rightarrow T \mathrm{e}^{\mathrm{i}x}, \quad \qquad \qquad \text{as } x \rightarrow +\infty.  
\end{cases}
\end{equation}

\subsection{Even-Odd Separation}
\label{ssec:Even-Odd Separation}

To solve this problem, we consider two cases where $f$ is either an even or odd function of $x$. Denoting these as $f_e$ and $f_o$ respectively, we can express $f$ as $f = f_e + f_o$. The equations for the even solution $f_e$ and the odd solution $f_o$ then become:
\begin{subequations}
\begin{align}
&\begin{cases}\label{eq:feequ}
\displaystyle f_e(x) = \int_{r}^{+ \infty} L_e(x, x') \left(\int^{+ \infty}_{r} K(x', x'') f_e(x'') dx''\right) \mathrm{d}x' , \quad\ \forall x > r,  \\
\displaystyle f_e(x) \rightarrow \frac{1}{2} (T+R) \mathrm{e}^{\mathrm{i}x} + \frac{1}{2} \mathrm{e}^{-\mathrm{i}x} , \quad\ \text{as } x \rightarrow \infty,  
\end{cases}
\\
&\begin{cases}\label{eq:foequ}
\displaystyle f_o(x) = \int_{r}^{+ \infty} L_o(x, x') \left(\int^{+ \infty}_{r} K(x', x'') f_o(x'') \: \mathrm{d} x''\right) \mathrm{d} x' , \quad \forall x > r,  \\
\displaystyle f_o(x) \rightarrow \frac{1}{2} (T-R) \mathrm{e}^{\mathrm{i}x} - \frac{1}{2} \mathrm{e}^{-\mathrm{i}x} , \quad \text{as } x \rightarrow \infty,  
\end{cases}
\end{align}
\end{subequations}
where dynamic kernel $K$ is defined in \eqref{eq:K}, and geometric kernel $L_e$, $L_o$ are defined by $S$, given in \eqref{eq:S}:
\begin{subequations}\label{eq:LeLo}
\begin{align}
L_e(x,x') &= S (x, x') + S (x, -x') \nonumber\\
&= - \upi^{-1} \left( \ln \left| \frac{x}{r} + \frac{r}{x} - \frac{x'}{r} - \frac{r}{x'} \right| + \ln \left| \frac{x}{r} + \frac{r}{x} + \frac{x'}{r} + \frac{r}{x'} \right| + 2 \gamma \right) , \\
L_o(x,x') &= S (x, x') - S (x, -x') \nonumber\\
&= - \upi^{-1} \left( \ln \left| \frac{x}{r} + \frac{r}{x} - \frac{x'}{r} - \frac{r}{x'} \right| - \ln \left| \frac{x}{r} + \frac{r}{x} + \frac{x'}{r} + \frac{r}{x'} \right| \right) . 
\end{align}
\end{subequations}

In the abstract operator language, \eqref{eq:feequ} can be written as
\begin{equation}
\begin{cases}\label{eq:feopeq}
\displaystyle f_e = \mathcal{L}_e \mathcal{K} f_e, \quad \forall x > r,  \\
\displaystyle f_e(x) \rightarrow \frac{1}{2} (T+R) \mathrm{e}^{\mathrm{i}x} + \frac{1}{2} \mathrm{e}^{-\mathrm{i}x} , \quad \text{as } x \rightarrow \infty,  
\end{cases}
\end{equation}
where $\mathcal{L}_e$ and $\mathcal{K}$ are the integral operators integrate from $r$ to $\infty$ with even geometric kernel $L_e$ given in \eqref{eq:LeLo} and dynamic kernel $K$ given in \eqref{eq:K}. And \eqref{eq:foequ} can be written as
\begin{equation}
\begin{cases}\label{eq:foopeq}
\displaystyle f_o = \mathcal{L}_o \mathcal{K} f_o, \quad \forall x > r,  \\
\displaystyle f_o(x) \rightarrow \frac{1}{2} (T-R) \mathrm{e}^{\mathrm{i}x} - \frac{1}{2} \mathrm{e}^{-\mathrm{i}x} , \quad \text{as } x \rightarrow \infty,  
\end{cases}
\end{equation}
where $\mathcal{L}_o$ is the integral operator with odd geometric kernel $L_o$ given in \eqref{eq:LeLo}. 

\subsection{Deviation from Far-Field Behavior}
\label{ssec:Deviation from Asymptotic Form}
If we define $f_e$ and $f_o$ in terms of their deviations from far fields by $h_e(x)$ and $h_o(x)$:
\begin{subequations}
\begin{align}\label{eq:feoexp}
f_e &= \frac{1}{2} (T+R) \mathrm{e}^{\mathrm{i}x} + \frac{1}{2} \mathrm{e}^{-\mathrm{i}x} + h_e(x), \\
f_o &= \frac{1}{2} (T-R) \mathrm{e}^{\mathrm{i}x} - \frac{1}{2} \mathrm{e}^{-\mathrm{i}x} + h_o(x),
\end{align}
\end{subequations}
We then derive equations for the near fields $h_e$ and $h_o$:
\begin{subequations}\label{eq:he+ho}
\begin{align}
&\begin{cases}\label{eq:heequ}
\displaystyle h_e = \mathcal{L}_e \mathcal{K} h_e + \frac{1}{2} (T+R) (\mathcal{L}_e \mathcal{K} - \mathcal{I}) \mathrm{e}^{\mathrm{i}x} + \frac{1}{2} (\mathcal{L}_e \mathcal{K} - \mathcal{I}) \mathrm{e}^{-\mathrm{i}x} , \quad \forall x > r,  \\
\displaystyle h_e(x) \rightarrow 0 , \quad \text{as } x \rightarrow \infty,  
\end{cases}
\\
&\begin{cases}\label{eq:hoequ}
\displaystyle h_o = \mathcal{L}_o \mathcal{K} h_o + \frac{1}{2} (T-R) (\mathcal{L}_o \mathcal{K} - \mathcal{I}) \mathrm{e}^{\mathrm{i}x} - \frac{1}{2} (\mathcal{L}_o \mathcal{K} - \mathcal{I}) \mathrm{e}^{-\mathrm{i}x} , \quad \forall x > r,  \\
\displaystyle h_o(x) \rightarrow 0 , \quad \text{as } x \rightarrow \infty,  
\end{cases}
\end{align}
\end{subequations}
where $\mathcal{I}$ is the identity operator, which takes a function to the same function. Based on the symmetry of \eqref{eq:he+ho}, we seek solutions of $h_e$ and $h_o$ in the form:
\begin{subequations}\label{eq:hepm+hopm}
\begin{align}
h_e &= \frac{1}{2} (T+R) h_{e, +} + \frac{1}{2} h_{e, -}, \label{eq:hepm} \\
h_o &= \frac{1}{2} (T-R) h_{o, +} - \frac{1}{2} h_{o, -}, \label{eq:hopm}
\end{align}
\end{subequations}
where $h_{e, \pm}$ and $h_{o, \pm}$ are to be determined. Substitute \eqref{eq:hepm+hopm} into \eqref{eq:he+ho} yields:
\begin{subequations}\label{eq:hepm+hopm_temp}
\begin{align}
(T + R) \{ h_{e,+} - \mathcal{L}_e \mathcal{K} h_{e,+} - (\mathcal{L}_e \mathcal{K} - \mathcal{I}) \mathrm{e}^{\mathrm{i}x} \} &= - \{ h_{e,-} - \mathcal{L}_e \mathcal{K} h_{e,-} - (\mathcal{L}_e \mathcal{K} - \mathcal{I}) \mathrm{e}^{- \mathrm{i}x} \} ,\label{eq:hepmequtemp}\\
(T - R) \{ h_{o,+} - \mathcal{L}_o \mathcal{K} h_{o,+} - (\mathcal{L}_o \mathcal{K} - \mathcal{I}) \mathrm{e}^{\mathrm{i}x} \} &= \{ h_{o,-} - \mathcal{L}_o \mathcal{K} h_{o,-} - (\mathcal{L}_o \mathcal{K} - \mathcal{I}) \mathrm{e}^{- \mathrm{i}x} \} .\label{eq:hopmequtemp}
\end{align}
\end{subequations}
A possible solution for $h_{e, \pm}$ and $h_{o, \pm}$ that satisfies \eqref{eq:hepm+hopm_temp} would be letting both sides equal to $0$:
\begin{equation}\label{eq:heopmequ}
h_{\chi, \pm} = \mathcal{L}_{\chi} \mathcal{K} h_{\chi,\pm} + (\mathcal{L}_{\chi} \mathcal{K} - \mathcal{I}) \mathrm{e}^{\pm \mathrm{i}x} ,
\end{equation}
where $\chi$ takes the values `$e$' for even and `$o$' for odd. $(\mathcal{L}_{\chi} \mathcal{K} - \mathcal{I}) \mathrm{e}^{\pm \mathrm{i}x}$ can be calculated analytically:
\begin{align}\label{eq:(LK-I)exp}
(\mathcal{L}_{\chi} \mathcal{K} - \mathcal{I}) \mathrm{e}^{\pm \mathrm{i}x} &=\int_{r}^{+ \infty} L_{\chi}(x, x') \left( \int_{r}^{+ \infty} K(x', x'') \mathrm{e}^{\pm \mathrm{i}x''} \: \mathrm{d} x'' \right) \mathrm{d} x' - \mathrm{e}^{\pm \mathrm{i}x} , \\
&= - \mathrm{e}^{\pm \mathrm{i}x} + \frac{1}{2 \sqrt{B}} \left[ - (\sqrt{B} \pm \mathrm{i}) \mathrm{e}^{-2\mathrm{i}\zeta} + (-\sqrt{B} \pm \mathrm{i}) \right] \mathrm{e}^{(\sqrt{B} \pm \mathrm{i}) r} I_{1, \chi} + I_{2, \chi, \pm} , 
\end{align}
where the expressions for $I_1$ and $I_2$ are given in Appendix \ref{app:expression}.

\subsection{Calculating $T$ \& $R$}

Finally, to ensure that the near fields $h_e$ and $h_o$ given in \eqref{eq:he+ho} approach to $0$ as $x \rightarrow \infty$, we have:
\begin{subequations}
\begin{align}
\frac{1}{2} (T+R) h_{e, +} + \frac{1}{2} h_{e, -} &\rightarrow 0, \quad \text{as } x \rightarrow \infty, \\
\frac{1}{2} (T-R) h_{o, +} - \frac{1}{2} h_{o, -} &\rightarrow 0, \quad \text{as } x \rightarrow \infty,
\end{align}
\end{subequations}
from which we can infer the coefficients $T$ and $R$:
\begin{subequations}\label{eq:T+R}
\begin{align}
T &= - \frac{1}{2} \left( C_e - C_o \right) ,\label{eq:T}\\
R &= - \frac{1}{2} \left( C_e + C_o \right) ,\label{eq:R}
\end{align}
where
\begin{equation}
    C_e = \lim_{x \rightarrow \infty} \frac{h_{e, -}}{h_{e, +}}, \qquad C_o = \lim_{x \rightarrow \infty} \frac{h_{o, -}}{h_{o, +}} .\label{eq:CeCo}
\end{equation}
\end{subequations}

Therefore, we have reduced the integral equation problem for $f(x) = \phi(x, 0)$ given in \eqref{eq:fequ} to solve the simpler integral equations in \eqref{eq:heopmequ}, and substitute into \eqref{eq:T+R} to obtain the transmission coefficient $T$ and reflection coefficient $R$. This approach guarantees that the solutions converge to the desired far-field behavior, ensuring the physical relevance and mathematical consistency of the results.

\subsection{Nyström method: Numerically Solving Integral Equations}
\label{ssec:Nyström method: Numerically Solving Integral Equations}

The integral equations in \eqref{eq:heopmequ} share the same form of
\begin{equation}
    h = \mathcal{L} \mathcal{K} h + g, \label{eq:int_ny}
\end{equation}
which can be effectively solved numerically using the Nyström method. This approach discretizes the integral operators $\mathcal{L}$ and $\mathcal{K}$ into matrices, transforming the integral equations into matrix equations. Specifically, the discretization is defined as:
\begin{equation}
    \boldsymbol{h}_i = h(x_i), \quad \boldsymbol{g}_i = g(x_i), \quad \mathsfbi{L}_{\chi, ij} = L_\chi(x_i, x_j) \Delta x, \quad \mathsfbi{K}_{ij} = K(x_i, x_j) \Delta x, 
\end{equation}
where $x_i = r + i\Delta x$, the step size $\Delta x = l / N$, and the integral range $l$ is truncated to ten times the wavelength, $20\upi$. The number of representative points, $N$, is selected as 2000 to ensure sufficient accuracy.

A convergence test was performed by halving and doubling $N$ for the relevant ranges of Bond number $B$ ($0.01 \leq B \leq 100$) and dimensionless radius $r$ ($0.01 \leq r \leq 10$), together with the range of slip parameter $\zeta$ ($0 \leq \zeta \leq \pi/2$). Among the tested combinations, the largest discrepancy of the calculated $T$ and $R$ arose at small values of $B=0.01$ and $r=0.01$, while $\zeta$ has a negligible influence on the overall error. The largest difference of the calculated $|T|$ between $N=1000$ and $N=2000$ reached approximately $3.6\%$. Increasing to $N=4000$ reduced the discrepancy of the calculated $|T|$ to about $0.8\%$. These results confirm that $N=2000$ is sufficiently accurate even under the most challenging scenario.

Special attention is given to the diagonal elements of $\mathsfbi{L}_{\chi, ij}$ due to the weak singularity at $x = x'$, arising from the logarithmic term $\ln | x/r + r/x - x'/r - r/x' |$. To address this, the singularity is mitigated using the following approximation:
\begin{equation}
    \int_{x_i}^{x_i+\Delta x} \ln \left| \frac{x}{r} + \frac{r}{x} - \frac{x_i}{r} - \frac{r}{x_i} \right| \: \mathrm{d} x \approx \ln \left( \left( \frac{1}{r} - \frac{r}{x_i^2} \right) \Delta x \right) - 1. 
\end{equation}
This procedure allows the integral equation \eqref{eq:int_ny} to be converted into the matrix equation:
\begin{equation}
    \boldsymbol{h} = \mathsfbi{L K} \boldsymbol{h} + \boldsymbol{g}, 
\end{equation}
which can be solved explicitly as:
\begin{equation}
    \boldsymbol{h} = ( \mathsfbi{I} - \mathsfbi{L K} )^{-1} \boldsymbol{g}, 
\end{equation}
where $\mathsfbi{I}$ denotes the identity matrix.

\section{Gravity Waves: Comparison with Previous Work}
\label{sec:Gravity Waves}

\begin{figure*}
    \centering   
    \includegraphics[width=0.8\textwidth]{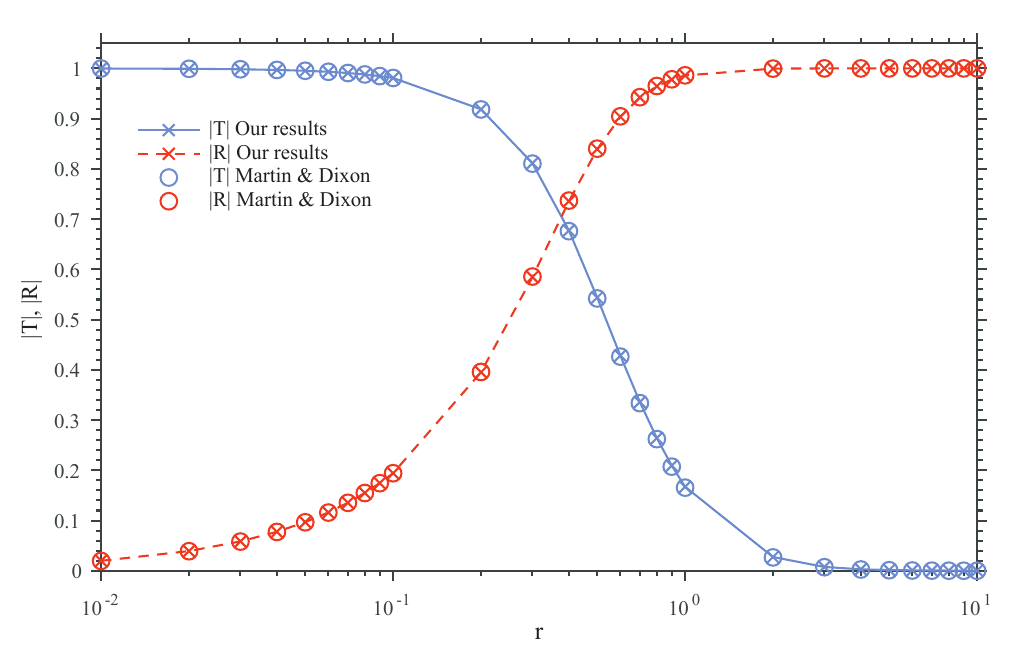}
    \caption{Comparison of results for gravity wave scattering between our semi-analytical results and results from \citet{martin1983scattering} for the transmission $|T|$ and reflection $|R|$ as a function of the dimensionless barrier radius $r$ (the dimensional radius $r'$ multiplied by the wavenumber $k$).}
    \label{fig:grav-compare}
\end{figure*}

For pure gravity waves ($B = \infty$), the second-order derivative term $d^2 \eta / dx^2$ in \eqref{eq:etaequ} becomes negligible. As a result, the boundary value problems (BVPs) for $\eta$ specified in these equations become self-contained solutions:
\begin{equation}
\begin{cases}\label{eq:etaequ_grav}
\displaystyle \eta = \phi \big|_{y=0}, & \forall x < -r,  \\
\displaystyle \eta \rightarrow \mathrm{e}^{\mathrm{i}x} + R \mathrm{e}^{-\mathrm{i}x}, & \text{as } x \rightarrow -\infty,
\end{cases}
\text{and }
\begin{cases}
\displaystyle \eta = \phi \big|_{y=0}, & \forall x > r,  \\
\displaystyle \eta \rightarrow T \mathrm{e}^{ \mathrm{i}x }, & \text{as } x \rightarrow +\infty.  
\end{cases}
\end{equation}

In conjunction with the solution to \eqref{eq:phiequ} provided in \eqref{eq:phi0sol}, we formulate the integral equation problem for $f(x) = \phi(x, 0)$:
\begin{equation}
\begin{cases}\label{eq:fequ_grav}
\displaystyle f(x) = \left( \int_{-\infty}^{-r} + \int_{r}^{+\infty} \right) \, S (x, x') f(x') \: \mathrm{d} x' ,  \quad \forall |x| > r,  \\
\displaystyle f(x) \rightarrow \mathrm{e}^{\mathrm{i}x} + R \mathrm{e}^{-\mathrm{i}x}, \qquad\ \text{as } x \rightarrow -\infty, \\
\displaystyle f(x) \rightarrow T \mathrm{e}^{\mathrm{i}x}, \quad \qquad \qquad \text{as } x \rightarrow +\infty,  
\end{cases}
\end{equation}
where $S$ is defined in \eqref{eq:S}. Following the analyses in sec.~\ref{ssec:Even-Odd Separation} and \ref{ssec:Deviation from Asymptotic Form}, we derive the equations for $h_{\chi, \pm}$ (where $\chi$ represents either even $h_{e, \pm}$ or odd $h_{o, \pm}$ solutions):
\begin{equation}\label{eq:heopmequ_grav}
h_{\chi, \pm} = \mathcal{L}_{\chi} h_{\chi,\pm} + (\mathcal{L}_{\chi} - \mathcal{I}) \mathrm{e}^{\pm \mathrm{i}x} ,
\end{equation}
where $\mathcal{L}_{\chi}$ (either $\mathcal{L}_{e}$ or $\mathcal{L}_{o}$) shares the kernel definition detailed in \eqref{eq:LeLo}. \eqref{eq:heopmequ_grav} requires numerical resolution, typically using the method outlined in sec.~\ref{ssec:Nyström method: Numerically Solving Integral Equations}. The resulting solutions $h_{e, \pm}$ and $h_{o, \pm}$ are then used to compute the transmission $T$ and reflection $R$ coefficients, as defined in Equations \eqref{eq:T+R}. The term $(\mathcal{L}_{\chi} - \mathcal{I}) \mathrm{e}^{\pm \mathrm{i}x}$ can be expressed analytically as:
\begin{equation}
        (\mathcal{L}_{\chi} - \mathcal{I}) \mathrm{e}^{\pm \mathrm{i}x} = I_{2, \chi, \pm} - \mathrm{e}^{\pm \mathrm{i}x} ,
\end{equation}
where $I_2$ is evaluated in Appendix~\ref{app:expression}. The semi-analytical results are compared with the multipole method results from \citet{martin1983scattering} in Fig.~\ref{fig:grav-compare} and Table~\ref{tb:grav-compare} in Appendix~\ref{app:gravity}, demonstrating their agreement on the reduction of transmission as the barrier size increases.

\section{Capillary-Gravity Waves: Phase Difference and Energy Conservation at Contact Line Limits}
\label{sec:No Dissipation}

In this section, we analyze the phase difference between the transmission and reflection coefficients, as well as the energy conservation of wave fields in two limiting cases of the contact line conditions: the fixed contact line ($\Lambda' = 0$) and the free contact line ($\Lambda' = \upi/2$). 

Recall \eqref{eq:T} and \eqref{eq:R}:
\begin{equation}\label{eq:TR_recall}
T = - \frac{1}{2} \left( C_e - C_o \right) , \qquad 
R = - \frac{1}{2} \left( C_e + C_o \right) ,
\end{equation}
where $C_\chi$ ($\chi$ takes the values `$e$' for even and `$o$' for odd) is given in \eqref{eq:CeCo} as
\begin{equation}
    C_\chi = \lim_{x \rightarrow \infty} \frac{h_{\chi, -}}{h_{\chi, +}}, \label{eq:C_chi_recall}
\end{equation}
and $h_{\chi, \pm}$ satisfies \eqref{eq:heopmequ}: 
\begin{equation}\label{eq:heopmequ_temp}
h_{\chi, \pm} = \mathcal{L}_{\chi} \mathcal{K} h_{\chi,\pm} + (\mathcal{L}_{\chi} \mathcal{K} - \mathcal{I}) \mathrm{e}^{\pm \mathrm{i}x},
\end{equation}
with the dynamic kernel $K$ given in \eqref{eq:K} and \eqref{eq:theta}. 

For the two limiting cases of contact lines, the dynamic kernel $K$ in \eqref{eq:K}  and \eqref{eq:theta} reduces to:
\begin{subequations}
    \begin{align}
    &\text{Fixed contact lines}~(\Lambda = 0): &K(x,x') &= \frac{1+B}{2\sqrt{B}} \left( - \mathrm{e}^{\sqrt{B}(2r - |x + x'|)} + \mathrm{e}^{-\sqrt{B}|x - x'|} \right), \\
    &\text{Free contact lines}~(\Lambda \to \infty): &K(x,x') &= \frac{1+B}{2\sqrt{B}} \left( \mathrm{e}^{\sqrt{B}(2r - |x + x'|)} + \mathrm{e}^{-\sqrt{B}|x - x'|} \right).
\end{align}
\end{subequations}
As such, $K$ is real values in both cases. Since the geometric kernels $L_\chi$ in \eqref{eq:LeLo} are also real from the outset, it follows that the complex conjugate of \eqref{eq:heopmequ_temp} is:
\begin{equation}
\label{eq:heopmequ_temp_cc}
h^*_{\chi, \pm} = \mathcal{L}_{\chi} \mathcal{K} h^*_{\chi,\pm} + (\mathcal{L}_{\chi} \mathcal{K} - \mathcal{I}) \mathrm{e}^{\mp \mathrm{i}x} .
\end{equation}
By comparing \eqref{eq:heopmequ_temp} with its complex conjugate \eqref{eq:heopmequ_temp_cc}, we find the symmetry:
\begin{equation}
    h^*_{\chi, \pm} = h_{\chi, \mp}. \label{eq:conj_sym}
\end{equation}
Using the symmetry $h^*_{\chi, -} = h_{\chi, +}$, \eqref{eq:C_chi_recall} reduces to:
\begin{equation}
    C_\chi = \lim_{x \rightarrow \infty} \frac{h_{\chi, -}}{h^*_{\chi, -}}, 
\end{equation}
which implies an unitary modular, $|C_\chi|=1$. We introduce a phase $\phi_\chi$ to rewrite $C_\chi$ as,
\begin{subequations}
\label{eq:C_chi_in_phi}
\begin{equation}
    C_\chi = \mathrm{e}^{2 \mathrm{i}\phi_\chi},
\end{equation}
and define the sum and difference between the phases of $C_e$ and $C_o$:
\begin{equation}
\phi_\mathrm{s} = \phi_e + \phi_o,~~~
\phi_\mathrm{d} = \phi_e - \phi_o.
\end{equation}
\end{subequations}
Substituting \eqref{eq:C_chi_in_phi} into \eqref{eq:TR_recall} yields the transmission coefficient $T$ and reflection coefficient $R$ expressed as:
\begin{subequations}\label{eq:TR_phi_sd}
\begin{align}
    T &= - \frac{\mathrm{e}^{2 \mathrm{i}\phi_e} - \mathrm{e}^{2 \mathrm{i}\phi_o}}{2} 
    = - \mathrm{e}^{ \mathrm{i}\phi_\mathrm{s}} \frac{\mathrm{e}^{ \mathrm{i}\phi_\mathrm{d} } - \mathrm{e}^{- \mathrm{i}\phi_\mathrm{d} }}{2}
     = - \mathrm{i} \mathrm{e}^{ \mathrm{i}\phi_\mathrm{s}} \sin\phi_\mathrm{d}      ,\label{eq:T_phi_sd} \\
    R &= - \frac{\mathrm{e}^{2 \mathrm{i}\phi_e} + \mathrm{e}^{2 \mathrm{i}\phi_o}}{2} 
    = - \mathrm{e}^{ \mathrm{i}\phi_\mathrm{s}} \frac{\mathrm{e}^{ \mathrm{i}\phi_\mathrm{d} } + \mathrm{e}^{- \mathrm{i}\phi_\mathrm{d} }}{2} 
    = - \mathrm{e}^{ \mathrm{i}\phi_\mathrm{s}} \cos\phi_\mathrm{d} ,
    \label{eq:R_phi_sd}
\end{align}
\end{subequations}
in which, $\phi_d$ determines the magnitudes of $T$ and $R$, and $\phi_s$ determines the phases of $T$ and $R$. Recall that $\phi_d$ and $\phi_s$ are associated with the near fields $h_e$ and $h_o$.

It then follows from \eqref{eq:TR_phi_sd} that:
\begin{equation}
    |T|^2 + |R|^2 = 1, \label{eq:nodiss}
\end{equation}
as no energy dissipation occurs in either the fixed or free contact line conditions.

It also follows from \eqref{eq:TR_phi_sd} that, although the phases of $T$ and $R$ individually depend on $\phi_s$ associated with the near fields, the phase difference between $T$ and $R$ does not rely on the near-field behaviors. The imaginary unit in \eqref{eq:T_phi_sd} reveals that the phase of the transmission coefficient $T$ has a $\upi/2$ phase difference compared to the reflection coefficient $R$ in \eqref{eq:R_phi_sd}:
\begin{equation}
    |\text{arg} \ T - \text{arg} \ R| = \frac{\upi}{2}. \label{eq:phasediff}
\end{equation}

\section{Capillary-Gravity Waves: Analytical Solution for an Infinitesimal Barrier}
\label{sec:Infinitesimal Barrier}

For the infinitesimal barrier case, the radius of the cylindrical barrier is assumed to satisfy $r \rightarrow 0$, implying that the barrier is much smaller than the characteristic wavelength of the incident waves. To maintain consistency with the linearized theory, this further requires that the surface displacement is sufficiently small, such that $k \eta_A \ll r \ll 1$. Under these assumptions, the liquid domain simplifies to the lower half-space $H$. Consequently, the boundary value problem (BVP) for $\phi$, initially defined in \eqref{eq:phiequ}, now transforms into:
\begin{equation}
\begin{cases}\label{eq:phiequ_infi}
\displaystyle \nabla^2 \phi = 0, & \forall (x, y) \in H, \\
\displaystyle \boldsymbol{\hat{n}} \bcdot \bnabla \phi = \eta, & \text{at } y = 0.
\end{cases}
\end{equation}
Similarly, the BVPs for $\eta$ \eqref{eq:etaequ} modify as follows:
\begin{subequations}
\begin{equation}
\begin{cases}\label{eq:etaequ1_infi}
\displaystyle \frac{\mathrm{d}^2 \eta}{\mathrm{d}x^2} - B \eta = - (1+B) \phi \big|_{y=0}, & \forall x < 0,  \\
\displaystyle \frac{\mathrm{d} \eta}{\mathrm{d}x} = \mathrm{i}\Lambda^{-1} \eta, & \text{at } x = 0, \\
\displaystyle \eta \rightarrow \mathrm{e}^{\mathrm{i}x} + R \mathrm{e}^{-\mathrm{i}x}, & \text{as } x \rightarrow -\infty,
\end{cases}
\end{equation}
and
\begin{equation}
\begin{cases}\label{eq:etaequ2_infi}
\displaystyle \frac{\mathrm{d}^2 \eta}{\mathrm{d}x^2} - B \eta = - (1+B) \phi \big|_{y=0}, & \forall x > 0,  \\
\displaystyle \frac{\mathrm{d} \eta}{\mathrm{d}x} = - \mathrm{i}\Lambda^{-1} \eta, & \text{at } x = 0, \\
\displaystyle \eta \rightarrow T \mathrm{e}^{ \mathrm{i}x }, & \text{as } x \rightarrow +\infty.  
\end{cases}
\end{equation}
\end{subequations}

Given that $H$ itself is the liquid domain, no transformation into a lower half-space is required. Thus, the conformal mapping $w = z/r + r/z$ in sec.~\ref{ssec:Conformal Mapping} simplifies to the identity $w = z$. Employing a similar approach as in sec.~\ref{ssec:Fourier Transform}, we formulate the integral equation problem for $f(x) = \phi(x, 0)$:
\begin{equation}
\begin{cases}\label{eq:fequ_infi}
\displaystyle f(x) = \int_{-\infty}^{+\infty} S (x, x') \eta(x') \: \mathrm{d} x' ,  \quad \forall x \in \mathbb{R},  \\
\displaystyle f(x) \rightarrow \mathrm{e}^{\mathrm{i}x} + R \mathrm{e}^{-\mathrm{i}x}, \qquad\ \text{as } x \rightarrow -\infty, \\
\displaystyle f(x) \rightarrow T \mathrm{e}^{\mathrm{i}x}, \quad \qquad \qquad \text{as } x \rightarrow +\infty,  
\end{cases}
\end{equation}
where $S$ is defined by:
\begin{equation}
S (x, x') = - \upi^{-1} \left( \ln{ \left| x - x' \right| } + \gamma \right). \label{eq:S_infi}
\end{equation}

Following the analysis in sec.~\ref{ssec:Even-Odd Separation} and \ref{ssec:Deviation from Asymptotic Form}, we derive the equations for $h_{\chi, \pm}$:
\begin{equation}\label{eq:heopmequ_infi}
h_{\chi, \pm} = \mathcal{L}_{\chi} \mathcal{K} h_{\chi,\pm} + (\mathcal{L}_{\chi} \mathcal{K} - \mathcal{I}) \mathrm{e}^{\pm \mathrm{i}x} ,
\end{equation}
where the kernels for $\mathcal{L}_{\chi}$ are redefined as follows:
\begin{subequations}
\begin{align}\label{eq:LeLo_infi}
L_e(x,x') &= S (x, x') + S (x, -x') \nonumber\\
&= - \upi^{-1} \left( \ln \left| x - x' \right| + \ln \left| x + x' \right| + 2 \gamma \right) , \\
L_o(x,x') &= S (x, x') - S (x, -x') \nonumber\\
&= - \upi^{-1} \left( \ln \left| x - x \right| - \ln \left| x + x' \right| \right) . 
\end{align}
\end{subequations}
and the kernel for $\mathcal{K}$ updates to:
\begin{equation}
K(x, x') = \frac{1+B}{2\sqrt{B}} \left( -\mathrm{e}^{-2\mathrm{i}\zeta} \mathrm{e}^{- \sqrt{B}|x + x'|} + \mathrm{e}^{-\sqrt{B}|x - x'|} \right). \label{eq:K_infi}
\end{equation}

The solution to \eqref{eq:heopmequ_infi} can be obtained using Fourier cosine and sine transforms. The detailed derivation and mathematical techniques are provided in Appendix~\ref{app:solution}. The derived solutions for $h_{e, \pm}$ and $h_{o, \pm}$ facilitate the calculation of transmission $T$ and reflection $R$ coefficients, as delineated in \eqref{eq:T+R}. These coefficients are expressed as:
\begin{subequations}\label{eq:T+R_infi}
\begin{align}
    T &= \frac{1}{2} \left( \frac{X-\mathrm{i}}{X+\mathrm{i}} + \frac{Y+\mathrm{i}}{Y-\mathrm{i}} \right) ,\label{eq:T_infi} \\
    R &= \frac{1}{2} \left( \frac{X-\mathrm{i}}{X+\mathrm{i}} - \frac{Y+\mathrm{i}}{Y-\mathrm{i}} \right) ,\label{eq:R_infi} 
\end{align}
where 
\begin{align}
    X &= \frac{1}{\upi} \left(\frac{3 + 2B}{\sqrt{3 + 4B}} \tan^{-1}\left(\sqrt{3 + 4B}\right) + \frac{1}{2} \ln(1 + B)\right) + \mathrm{i}\frac{(3 + B) \tan\zeta}{2 \sqrt{B}} , \\
    Y &= \frac{1}{\upi} \left(\frac{3 + 6B + 2B^2}{\sqrt{3 + 4B}} \tan^{-1}\left(\sqrt{3 + 4B}\right) - \frac{1}{2} \ln(1 + B)\right) - \mathrm{i}\frac{(3 + B) \sqrt{B}}{2 \tan\zeta} .
\end{align}
\end{subequations}
This solution expresses $T$ and $R$ as functions of the dimensionless Bond number $B$ defined in \eqref{eq:B} and dimensionless slip parameter $\zeta$ defined in \eqref{eq:theta}.

In Hocking's work \citep{ref:Hocking-1987} considering scattering from an infinitesimally thin barrier of a finite immersed depth, an analytical solution arises in the limit of a zero immersed depth $d \rightarrow 0$. In contrast, our solution is derived from a cylindrical barrier geometry, reducing to an infinitesimal barrier as the radius $r \rightarrow 0$. In Appendix~\ref{app:hocking}, we show that the two solutions are equivalent even though they approach an infinitesimal barrier in different ways. Our theory underscores a generalization from an idealized infinitesimally thin barrier to a cylindrical barrier of practical applicability.

\subsection{Fixed Contact Line}

For the case of infinitesimal barrier ($r \rightarrow 0$) with a fixed contact line ($\zeta = 0$), \eqref{eq:T+R_infi} reduce to:
\begin{subequations}\label{eq:T+R_infi_fixed}
\begin{align}
    T &= \frac{1}{2} \left( \frac{X-\mathrm{i}}{X+\mathrm{i}} + 1 \right) 
    =\frac{X}{X+\mathrm{i}},\label{eq:T_infi_fixed} \\
    R &= \frac{1}{2} \left( \frac{X-\mathrm{i}}{X+\mathrm{i}} - 1 \right) 
    = - \frac{\mathrm{i}}{X+\mathrm{i}} 
    ,\label{eq:R_infi_fixed}
\end{align}
where
\begin{equation}
    X = \frac{1}{\upi} \left(\frac{3 + 2B}{\sqrt{3 + 4B}} \tan^{-1}\left(\sqrt{3 + 4B}\right) + \frac{1}{2} \ln(1 + B)\right) .
\end{equation}
\end{subequations}
Here $R$ and $T$ satisfies the $90^\circ$ phase difference revealed in the prior section.

It follows from \eqref{eq:T+R_infi_fixed} that 
\begin{equation}
    R = T - 1, 
\end{equation}
implying that the diffracted field, $\phi_D = \phi - \phi_I$, is symmetric about the barrier, that is, 
\begin{equation}\label{eq:infi_fixed_sym}
\phi_D(x, y) = \phi_D(-x, y).    
\end{equation}

\subsection{Free Contact Line}

For the case of infinitesimal barrier ($r \rightarrow 0$) with a free contact line ($\zeta = \upi/2$), \eqref{eq:T+R_infi} reduce to:
\begin{subequations}\label{eq:T+R_infi_free}
\begin{align}
    T &= \frac{1}{2} \left( 1 + \frac{Y+\mathrm{i}}{Y-\mathrm{i}} \right)
    = \frac{Y}{Y-\mathrm{i}},\label{eq:T_infi_free} \\
    R &= \frac{1}{2} \left( 1 - \frac{Y+\mathrm{i}}{Y-\mathrm{i}} \right)
    = \frac{\mathrm{i}}{Y-\mathrm{i}} ,\label{eq:R_infi_free} 
\end{align}
where 
\begin{equation}
    Y = \frac{1}{\upi} \left(\frac{3 + 6B + 2B^2}{\sqrt{3 + 4B}} \tan^{-1}\left(\sqrt{3 + 4B}\right) - \frac{1}{2} \ln(1 + B)\right) ,
\end{equation}
\end{subequations}
Again, $R$ and $T$ satisfies the $90^\circ$ phase difference revealed in the prior section.

It follows from \eqref{eq:T+R_infi_free} that 
\begin{equation}
    -R = T - 1 ,
\end{equation}
implying that the diffracted field, $\phi_D = \phi - \phi_I$, is anti-symmetric about the barrier, that is,
\begin{equation}\label{eq:infi_free_sym}
\phi_D(x, y) = - \phi_D(-x, y).    
\end{equation}

\section{From an Infinitesimal Barrier to a Barrier with a Finite Cross-Section: No Dissipation}
\label{sec:Capillary-Gravity Waves without Dissipation}

\begin{figure*}
    \centering   
    \includegraphics[width=1\textwidth]{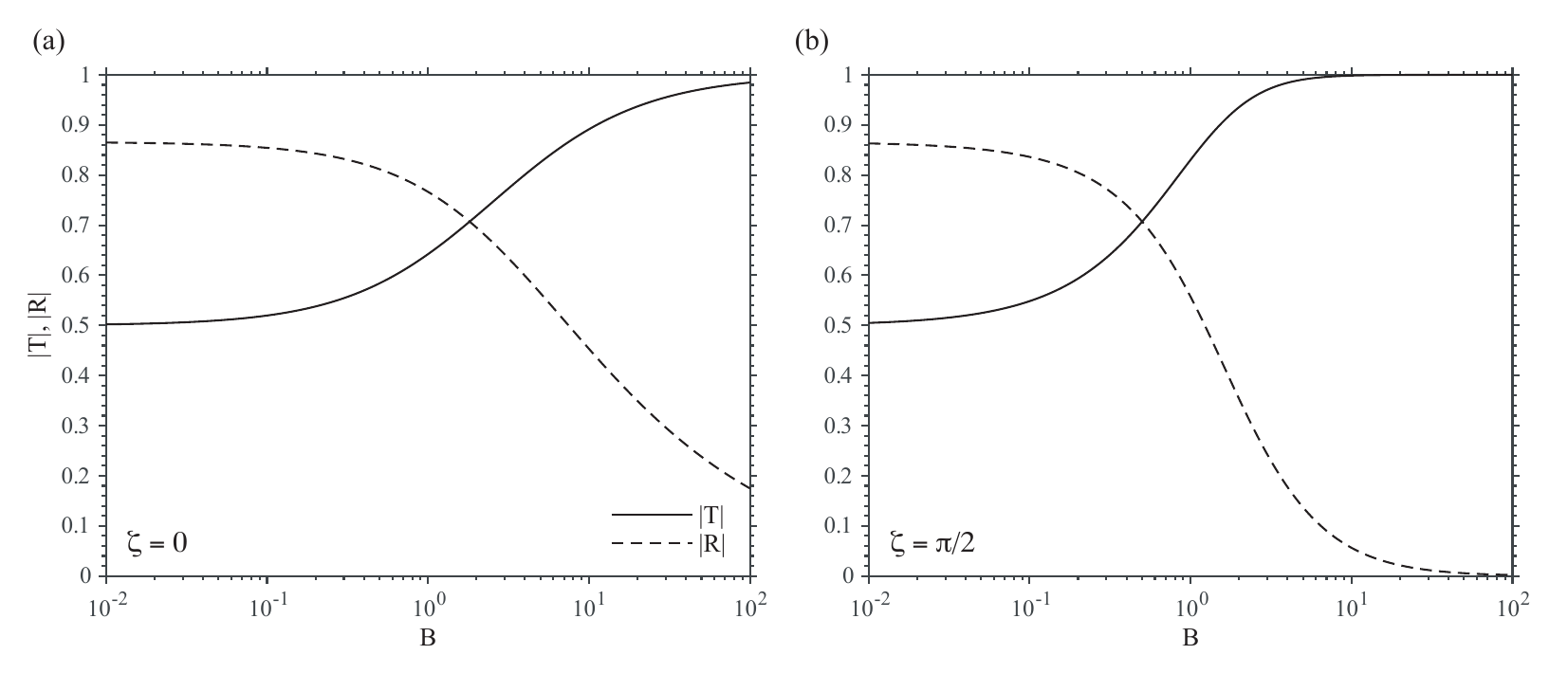}
    \caption{Transmission $|T|$ and reflection $|R|$ as functions of the Bond number $B$ (defined in (\ref{eq:B})) when scattered by an infinitesimal barrier $r=0$: (a) fixed and (b) free contact lines, following from \eqref{eq:T+R_infi_fixed} and \eqref{eq:T+R_infi_free}, respectively.}
    \label{fig:result1}
\end{figure*}

The transmission properties of surface waves interacting with an infinitesimal barrier, as calculated from the closed-form solution \eqref{eq:T+R_infi_fixed} and \eqref{eq:T+R_infi_free}, are shown in Fig.~\ref{fig:result1}, revealing distinct behaviors depending on the Bond number. Fig.~\ref{fig:result1}(a) presents the magnitudes of transmission $|T|$ and reflection $|R|$ as functions of the Bond number $B$ with a fixed contact line ($\zeta = 0$). In the capillary limit ($B \ll 1$), the transmission coefficient asymptotically approaches $|T| = 0.5$, indicating that only 25\% of the capillary wave energy is transmitted, while the remaining 75\% energy driven by surface tension is reflected by the fixed contact lines. As the system transitions to the gravity-dominated regime ($B \gg 1$), the influence of the contact line diminishes, and the transmission approaches $|T| = 1$. In this regime, the barrier no longer significantly affects the wave, allowing the full wave energy to pass through. 

Fig.~\ref{fig:result1}(b) shows the results for a free contact line ($\zeta = \upi/2$). The behavior in the capillary and gravity limits ($B \ll 1$ and $B \gg 1$) are similar to the fixed contact line case, with the transmission coefficient approaching $|T| = 0.5$ and 1, respectively. However, in the intermediate Bond number range, the free contact lines allow for the transmission greater than that of the fixed contact lines, as depicted by a comparison between Fig.~\ref{fig:result1}(a) and (b). Such a transmission difference between the two cases follows from different coupling of the energy from the incident side to the transmitted side, as evident in the symmetry difference of the diffracted fields between the two cases, as noted earlier in \eqref{eq:infi_fixed_sym} versus \eqref{eq:infi_free_sym}.

\begin{figure*}
    \centering   
    \includegraphics[width=1\textwidth]{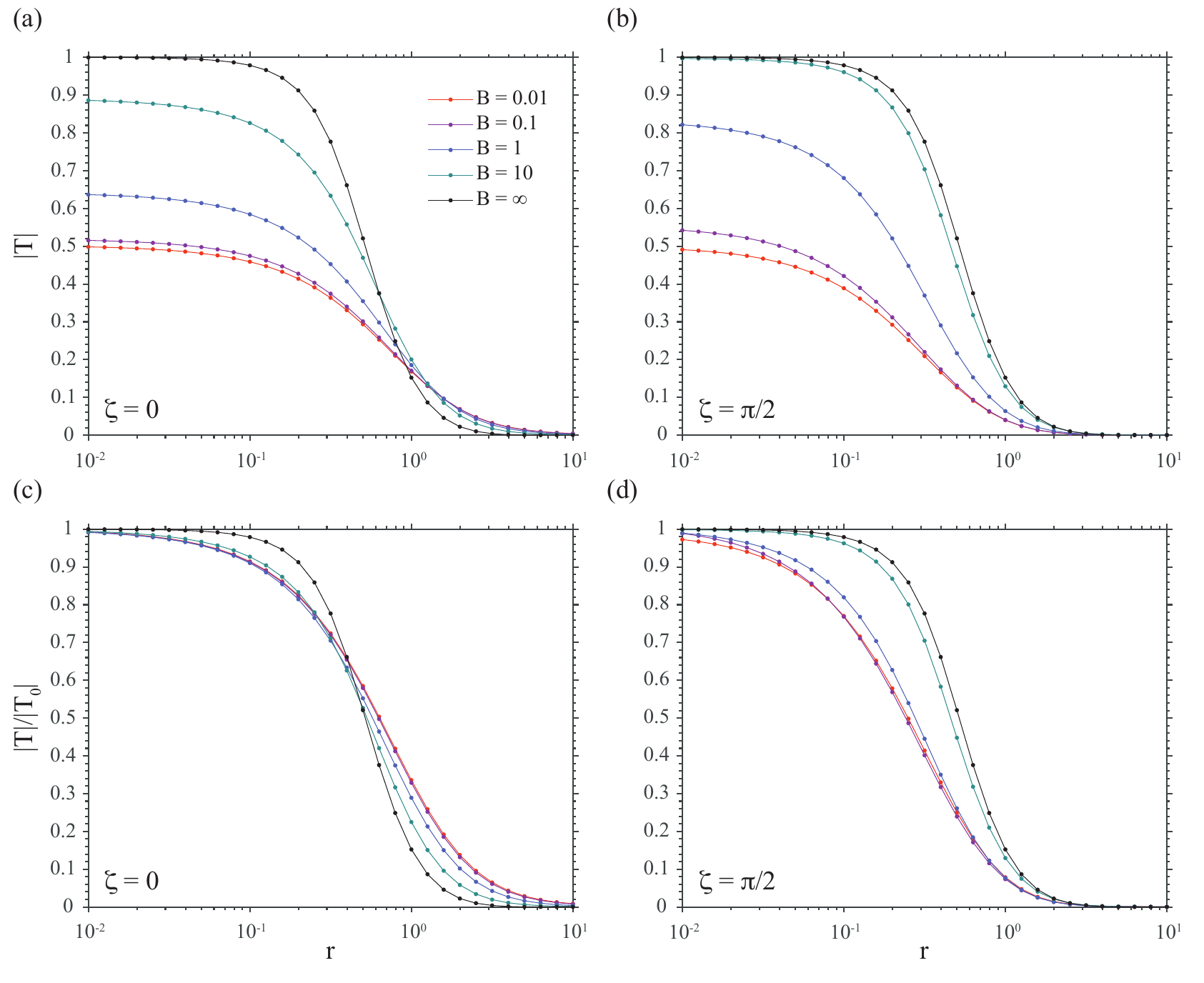}
    \caption{(a) and (b): Transmission coefficient $|T|$ as a function of the dimensionless barrier radius $r$ for different Bond number values for fixed and free contact lines, $\zeta = 0$ and $\pi/2$, respectively. (c) and (d) are the same with (a) and (b), respectively, but with the transmission coefficient $|T|$ normalized by the transmission in the infinitesimal barrier limit $|T_0|$.}
    \label{fig:result2}
\end{figure*}

As the barrier is increased from zero size to a finite cylindrical geometry, the transmission depends on the Bond number and barrier size. For a range of Bond number $B$ values, the transmission coefficient is calculated from the semi-analytical method using \eqref{eq:T} and illustrated in Fig.~\ref{fig:result2}(a) and (b) for dependence on the dimensionless barrier radius ($r=kr'$). In the limit of small barrier sizes ($r \ll 1$), the transmission $|T|$ approaches values ranging from 0.5 to 1 when varying the Bond number from capillary-dominant regime ($B \ll 1$) to gravity-dominant regime ($B \gg 1$), as noted earlier in Fig.~\ref{fig:result1}. As the barrier radius becomes larger ($r \gg 1$), the transmission coefficient across different Bond number values uniformly decreases due to the barrier’s finite cross-section. Compared to the fixed contact line, the overall transmission $|T|$ is higher for all $B$ under the free contact line. 

There is an interplay of Bond number and barrier size. For the fixed contact lines [Fig.~\ref{fig:result2}(a)], the transmission showcases a contrast when the radius $r$ crosses around 1: at small $r$, higher $B$ leads to higher transmission, but at large $r$, lower $B$ results in higher transmission. In contrast, for the free contact lines [Fig.~\ref{fig:result2}(b)], the barrier size effect appears uniform across different $B$: the higher the $B$, the higher the transmission. If we normalize the transmission $|T|$ by the infinitesimal-barrier limit $|T_0|$ (from the closed-form solution \eqref{eq:T_infi}), the ratio $|T| / |T_0|$ highlights the size effect across different Bond numbers.

The influence of barrier size as examined by the normalized transmission $|T|/|T_0|$ is now shown in Fig.~\ref{fig:result2}(c) and (d). The results show that for small barrier radii ($r<0.1$), the effect of barrier size is minimal, reducing transmission by no more than 10\%. However, as the barrier radius becomes comparable to the wavelength ($0.1 < r < 1$), transmission decreases substantially, pointing to a more significant interaction between the wave and the barrier. For large barriers  ($r > 1$), transmission is reduced to less than 20\%, indicating a dramatic size effect. These findings demonstrate a clear size-dependent transmission behavior, regardless of the Bond number. All key observations in this section have been summarized in Table~\ref{tab:summary}.

\begin{table}
\centering
\caption{Summary of transmission behaviors for various Bond numbers ($B$) and barrier radii ($r$) at the cases of fixed and free contact lines.}
\label{tab:summary}
\begin{tabularx}{\linewidth}{l|X}
\hline
\textbf{Condition / Parameter} 
& \textbf{Key Observations} \\
\hline

\textbf{Infinitesimal Barrier ($r=0$)} 
& \begin{itemize}[leftmargin=0pt]
    \item \textit{Fixed contact line} ($\zeta = 0$):
    \begin{itemize}
        \item Analytical solution \eqref{eq:T+R_infi_fixed}.
        \item Symmetry of the diffracted field: $T-1=R$
        \item Capillary wave limit $B \ll 1$: $\lvert T \rvert \to 0.5$ (25\% of energy transmitted).
        \item Gravity wave limit $B \gg 1$: $\lvert T \rvert \to 1$ (full transmission).
    \end{itemize}
    \item \textit{Free contact line} ($\zeta = \pi/2$):
    \begin{itemize}
        \item Analytical solution \eqref{eq:T+R_infi_free}.
        \item Anti-symmetry of the diffracted field: $T-1=-R$
        \item Similar to fixed contact lines in the extreme $B$ limits.
        \item Intermediate $B$: Higher $\lvert T\rvert$ than for fixed contact lines.
    \end{itemize}

    \item \textit{Phase difference}: $|\text{arg} \ T - \text{arg} \ R| = \upi/2$.

    \item Results presented in Fig.~\ref{fig:result1}.
\end{itemize} 
\\ \hline

\textbf{Barrier with Nonzero Radius ($r>0$)} 
& \begin{itemize}[leftmargin=0pt]
    \item \textit{Small radius} ($r \ll 1$): Behavior matches the infinitesimal-barrier limit (varying from $0.5$ to $1$, depending on $B$).
    \item \textit{Intermediate radius} ($r \sim 1$): Noticeable decrease in $\lvert T\rvert$ due to stronger wave-barrier interaction.
    \item \textit{Large radius} ($r \gg 1$): $\lvert T\rvert$ can fall below $20\%$ of the infinitesimal-barrier value.
    \item Free contact line ($\zeta = \pi/2$) generally yields higher $\lvert T\rvert$ than fixed contact line ($\zeta = 0$).
    
    \item Results presented in Fig.~\ref{fig:result2}(a-b).
\end{itemize} 
\\ \hline

\textbf{Normalized Transmission} 
& \begin{itemize}[leftmargin=0pt]
    \item $\lvert T_0\rvert =$ Transmission for $r=0$ (infinitesimal barrier).
    \item For $r < 0.1$: $|T| \approx |T_0|$ (less than 10\% difference).
    \item For $r > 1$: $|T|$ often $< 20\%$ of $|T_0|$.
    
    \item Results presented in Fig.~\ref{fig:result2}(c-d).
\end{itemize}
\\
\hline
\end{tabularx}
\end{table}

\section{Dissipation due to Dynamic Contact Lines}
\label{sec:dissipation}

\begin{figure*}
    \centering   
    \includegraphics[width=1\textwidth]{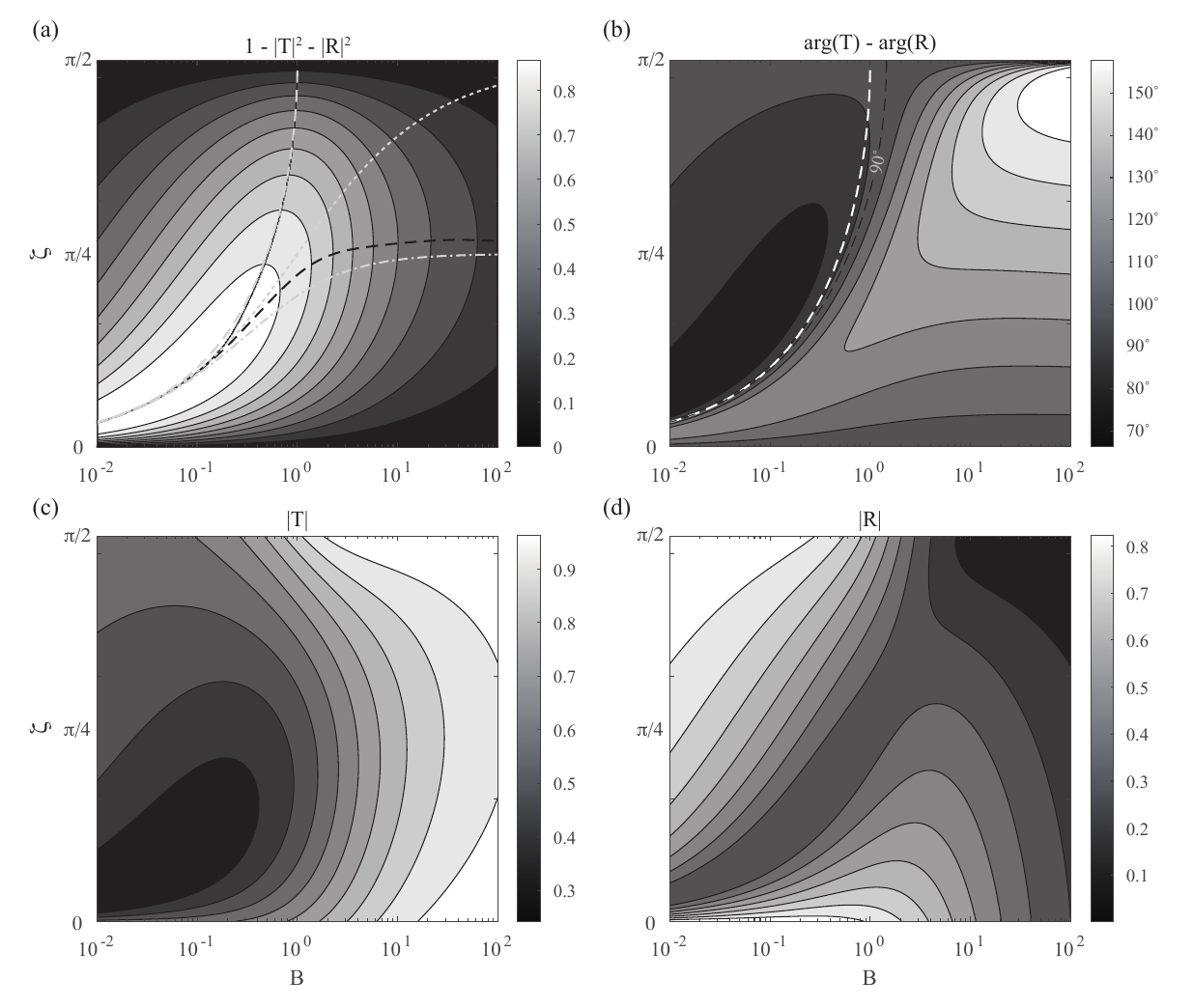}
    \caption{Intensity plots of (a) energy dissipation $1 - |T|^2 - |R|^2$, (b) phase difference $\arg T - \arg R$, (c) transmission $|T|$, and (d) reflection $|R|$ in the parameter space of the Bond number $B$ and the slip parameter $\zeta$ (defined in \eqref{eq:theta}) for scattering in the infinitesimal barrier limit, following from \eqref{eq:T+R_infi}. The black dashed line represents the slip coefficient values where maximum dissipation occurs at given $B$, and is compared with the white dash-dotted line and the white dotted line, representing the slip coefficient equal to the phase speed of capillary waves from \eqref{eq:c_c} and the phase speed of capillary-gravity waves from \eqref{eq:c_cg}, respectively. The black solid line indicates the ridge of dissipation, representing the path of the least descent for the dissipation in the parameter space of $B$ and $\zeta$, and coincides with the white dashed line from \eqref{eq:c_?}.}
    \label{fig:diss_infi}
\end{figure*}

For general contact line dynamics with a finite slip coefficient $\Lambda'$, energy dissipation occurs near the contact lines due to the slipping motion. In our model, which analyzes the scattering of capillary-gravity waves by a fixed, semi-immersed cylindrical barrier with dynamic contact lines, this dissipation depends on the dimensionless barrier radius $r$, the Bond number $B$, and the slip parameter $\zeta$. We are interested on finding the slip coefficient values where the dissipation is most effective.

For an infinitesimal barrier ($r \ll 1$), the results are presented in the parameter space of $B$ and $\zeta$ (Fig.~\ref{fig:diss_infi}) in terms of partition of energy dissipation $E_s = 1 - |T|^2 - |R|^2$, phase relation $\arg T - \arg R$, transmission $|T|$, and reflection $|R|$, calculated from \eqref{eq:T+R_infi}. In Fig.~\ref{fig:diss_infi}, the slip coefficient for maximum dissipation at a given Bond number $B$ is shown by the black dashed line. We also include two reference lines: the condition where the slip coefficient $\Lambda'$ equals the phase speed of pure capillary waves  (white dash-dotted line),
\begin{subequations}\label{eq:c_c}
\begin{equation}
\Lambda' = c_c = \sqrt{\sigma k / \rho},
\end{equation} 
corresponding to 
\begin{equation}
\tan\zeta = \sqrt{B/(1+B)};
\end{equation}
\end{subequations}
and the condition where the slip coefficient $\Lambda'$ equals phase speed of capillary-gravity waves (white dotted line),
\begin{subequations}\label{eq:c_cg}
\begin{equation}
\Lambda' = c_p = \sqrt{\sigma k / \rho} \cdot \sqrt{1+B},
\end{equation} 
corresponding to 
\begin{equation}
\tan\zeta = \sqrt{B}.
\end{equation}
\end{subequations}

For small $B$, the two reference lines align closely with the calculated maximum dissipation line. However, as $B$ increases, the maximum dissipation line deviates from both reference lines but remains closer to the $\Lambda' = c_c$ line. This behavior can be interpreted by wave impedance matching with the impedance of the contact line motion. The results suggest that the incident energy is more effectively coupled to the contact line boundary when $\Lambda'$ is closer to the phase speed of the pure capillary wave ($c_c$) than the capillary-gravity wave ($c_p$). This proximity to $\Lambda' = c_c$ reflects the dominant role of surface tension effects in determining energy dissipation at any Bond numbers. The results extend a finding in \citet{zhang2013} that, in the limit of pure capillary waves ($B \rightarrow 0$) and infinitesimal barrier ($r\rightarrow 0$), maximum dissipation occurs when the slip coefficient $\Lambda'$ equals the phase speed of the incident capillary waves. Extending this analysis, we examine energy dissipation beyond pure capillary waves, incorporating an arbitrary Bond number $B$.

Also shown in Fig.~\ref{fig:diss_infi}(a) is the ridge of dissipation (black solid line), following from the minima of the magnitude of the gradients of the dissipation in the parameter space. The line represents the path of least descent for dissipation in the parameter space. Intriguingly, it is observed that this dissipation ridge line fits with the relation (white dashed line)
\begin{subequations}\label{eq:c_?}
\begin{equation}
\Lambda' = \sqrt{\sigma k / \rho} \cdot \sqrt{(1+B)/(1-B)},    
\end{equation}
corresponding to
\begin{equation}
\tan\zeta = \sqrt{B/(1-B)}.
\end{equation}
\end{subequations}
The physical interpretation of this coincidence remains unclear and warrants further investigation. The phase difference $\arg T - \arg R$ depicted in Fig.~\ref{fig:diss_infi}(b), generally differing from $90^\circ$ with the dissipation, crosses $90^\circ$ around the ridge of the dissipation.

\begin{figure*}
    \centering   
    \includegraphics[width=1\textwidth]{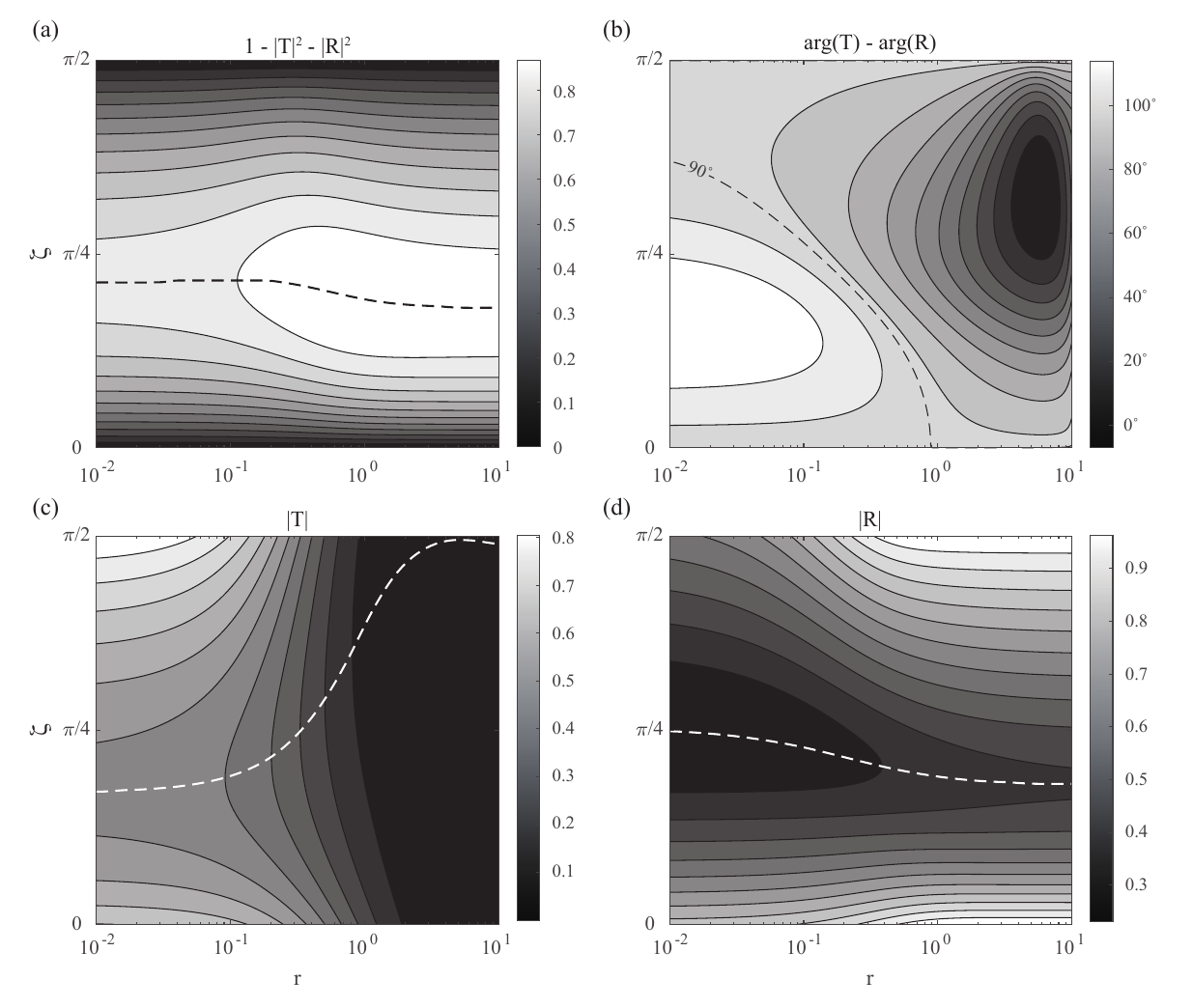}
    \caption{Intensity plots of (a) $1 - |T|^2 - |R|^2$, (b) $\arg T - \arg R$, (c) $|T|$, and (d) $|R|$ in the parameter space of the dimensionless barrier radius $r$ and slip parameter $\zeta$ for a Bond number $B = 1$. The dashed lines correspond to (a) a maximum $1 - |T|^2 - |R|^2$, (b) a $90^\circ$ phase difference, (c) a minimum $|T|$, and (d) a minimum $|R|$ at given $B$.}
    \label{fig:diss_fi}
\end{figure*}

As the barrier size increases from zero to a finite value, the calculated results of the dissipation partition $E_s$, phase difference $\arg T - \arg R$, transmission $|T|$, and reflection $|R|$ are presented in Fig.~\ref{fig:diss_fi} in the parameter space of $r$ and $\zeta$ for a Bond number $B = 1$, a typical value when surface tension and gravity effects are comparable. The results of the dissipation partition in Fig.~\ref{fig:diss_fi}(a) show that contact line dissipation remains significant even as the barrier size increases. The slip parameter $\zeta$ corresponding to maximum dissipation for a given $r$ stays near $0.2\upi$ across the range of $r$, where the dissipation remains approximately $80\%$. Increasing the barrier size has little effect on the overall energy dissipation, as the dissipation is due to the contact lines and is largely independent of the barrier size. Nevertheless, the phase difference is sensitive to the size, as depicted in Fig.~\ref{fig:diss_fi}(b).

Even though the barrier size has little effect on the overall energy dissipation, the transmission and reflection individually have a strong dependence on the size. The transmission shown in Fig.~\ref{fig:diss_fi}(c) highlights a significant reduction as the barrier radius increases, particularly for $r > 1$, where the barrier size becomes comparable to the wavelength for the transmission to diminish to nearly negligible levels. This sharp decrease underscores the dominance of reflection and/or dissipation when the barrier size exceeds the wavelength scale. Also shown is the slip parameter $\zeta$ for the minimum transmission for a given $r$ (white dashed line), which changes from $0.17 \upi$ to $0.5 \upi$, with particularly rapid variation near $r = 1$. 

The reflection, depicted in Fig.~\ref{fig:diss_fi}(d), generally increases with increasing size $r$. The value of $\zeta$ corresponding to minimum reflection for a given $r$ gradually shifts from $0.25 \upi$ to $0.18 \upi$.  This indicates that the reflection is less sensitive to changes in barrier size compared to transmission, which could imply a relatively stable reflection behavior over a range of barrier sizes.

Overall, results from Fig.~\ref{fig:diss_fi} suggest that while the transmission is heavily affected by increasing barrier size, dissipation and reflection exhibit more complex but less drastic changes. The interplay between $r$ and $\zeta$ across the different plots provides a comprehensive view of the wave dynamics influenced by the barrier. All key observations in this section have been summarized in Table~\ref{tab:dissipation_summary}.

\begin{table}
\centering
\caption{Summary of dissipation properties and wave transmission/reflection for dynamic contact lines. 
Parameters: Bond number $B$, slip parameter $\zeta$, and barrier radius $r$.}
\label{tab:dissipation_summary}
\begin{tabularx}{\linewidth}{l|X}
\hline
\textbf{Parameter / Condition} & \textbf{Key Observations} \\
\hline
\textbf{Infinitesimal Barrier ($r \ll 1$)} 
& \begin{itemize}[leftmargin=0pt]
    \item Analytical solution \eqref{eq:T+R_infi}.
    \item \textit{Maximum Dissipation Line} occurs near slip speed $\Lambda' \approx c_c$ for small $B$, deviating moderately for larger $B$, where $c_c = \sqrt{\sigma k /\rho}$ is the phase speed of capillary waves.
    \item Maximum dissipation line is \textit{closer} to $c_c$ than to $c_p$ for a broad range of $B$, indicating the importance of surface tension effects, where $c_p = \sqrt{\sigma k / \rho}\sqrt{1+B}$ is the phase speed of capillary-gravity waves.
    \item \textit{Dissipation Ridge} follows $\Lambda' \approx \sqrt{\sigma k/\rho}\sqrt{\tfrac{1+B}{1-B}}$; its exact physical significance needs further study.
    \item \textit{Phase Difference}, $\arg T - \arg R$, crosses $90^\circ$ near the dissipation ridge.
    \item Results presented in Fig.~\ref{fig:diss_infi}.
\end{itemize} \\
\hline
\textbf{Barrier with Nonzero Radius ($r>0$)} 
& \begin{itemize}[leftmargin=0pt]
    \item \textit{Overall Dissipation} remains high (e.g., $\sim 80\%$ at $B=1$) and is not strongly affected by increasing $r$.
    \item \textit{Transmission} $|T|$ drastically decreases for $r \gtrsim 1$, reflecting stronger blockage.
    \item \textit{Reflection} $|R|$ generally increases as $r$ increases, though less sharply than $|T|$ decreases.
    \item \textit{Slip Coefficient for Maximum Dissipation} remains near $\zeta \approx 0.2\upi$ across a range of $r$ (results for $B=1$).
    \item Results presented in Fig.~\ref{fig:diss_fi}.
    
\end{itemize}
\\ \hline
\textbf{Implications} 
& \begin{itemize}[leftmargin=0pt]
    \item Matching the contact-line slip coefficient $\Lambda'$ near capillary-wave speed leads to significant energy dissipation.
    \item Barrier size modifies transmission/reflection more than total dissipation.
\end{itemize} 
\\ \hline
\end{tabularx}
\end{table}

\section{Summary and discussion}
\label{sec:Summary}

This study develops a comprehensive semi-analytical framework to investigate the scattering of capillary-gravity waves by a fixed, semi-immersed cylindrical barrier while incorporating the dynamic contact line condition. The formulation begins with the linearized potential flow assumption, leveraging an effective-slip model to connect the contact line velocity and the deviation of the contact angle from its equilibrium state. This approach extends classical theories of pure gravity wave scattering by a thin barrier to include the effect of surface tension, dynamic contact lines, and a cylindrical barrier geometry with nonzero cross-section, thereby providing a more complete representation of realistic wave-barrier interactions.

Two coupled boundary value problems regarding the velocity potential and surface elevation, formulated from the governing equations and boundary conditions, are solved using techniques such as differential equations, conformal mapping, and Fourier transforms.  These problems are combined into a compound integral equation, which is simplified using advanced integral equation techniques and then numerically solved via discretization with the Nyström method. Numerous integrals were analytically resolved using specialized techniques to streamline the formulation. 

Benchmarking against prior pure-gravity results validates the consistency and accuracy of the theoretical framework. Under the limiting cases of fixed and free contact line conditions, our analysis of capillary-gravity wave scattering revisits the fundamental energy-conserving behaviors and reveals a $90^\circ$ phase difference between the transmission and reflection coefficients, $T$ and $R$.

A key theoretical contribution of our analysis of capillary-gravity wave scattering is the closed-form solution \eqref{eq:T+R_infi} derived for an infinitesimal barrier using Fourier cosine and sine transforms, along with advanced integral equation techniques. This solution reveals that the diffracted field is symmetric or anti-symmetric between the incident and transmitted sides under fixed or free contact line conditions of no dissipation, respectively. The analysis for no dissipation also characterizes reduced transmission ($|T_0|$) varying from 1 to 0.5 as the waves transition from gravity waves to capillary waves. This analytical result then serves as a baseline for illustrating how scattering characteristics evolve from an infinitesimal barrier radius to larger barrier radii using the ratio $|T|/|T_0|$, providing insights into the transition from nearly unimpeded wave propagation ($|T|/|T_0|=1$) to regimes where the barrier significantly alters the transmission.

Our analysis of energy dissipation due to dynamic contact lines reveals the effects of barrier radius, Bond number, and slip coefficient in capillary-gravity wave scattering. For an infinitesimal barrier, the maximum dissipation occurs when the slip coefficient approximates the phase speed of capillary waves given by \eqref{eq:c_c}, supporting previous findings for impedance matching between pure capillary waves and contact line motions while extending the analysis to arbitrary Bond numbers. In the parameter space of Bond number and slip parameter, the dissipation ridge, representing the path of least descent, was found to coincide with a special relation given in \eqref{eq:c_?}, where the phase difference crosses 90$^\circ$, suggesting certain physics behind that requires further investigation. For a barrier with a nonzero radius, dissipation remains significant and is largely independent of barrier size, maintaining approximately $80\%$ as the maximum dissipation across all barrier sizes. Transmission decreases sharply as the barrier size becomes comparable to the wavelength, while reflection increases moderately with barrier size, showing less sensitivity to barrier size than transmission. These findings broaden the theoretical understanding of fluid–structure interactions under the influence of surface tension and dynamic contact line conditions. 

Despite these insights, our framework has several limitations. First, we adopt a linearized potential flow model, so strong nonlinear effects—such as wave breaking, steepening, or large-amplitude deformations—fall outside the scope of our analysis. Viscous mechanisms are treated only implicitly through an effective-slip parameter, and meniscus phenomena near the contact line are not captured in detail. Moreover, the barrier is assumed to be semi-immersed in infinite (or effectively deep) water, which simplifies the geometry to a simply connected domain amenable to standard conformal mapping. Relaxing these assumptions to account for fully submerged barriers, finite-depth conditions, or more general body shapes requires additional mathematical or numerical developments.

Several directions arise naturally for future work:

(a) \emph{Barrier position relative to the free surface.}
What if the cylindrical barrier is not semi-immersed (e.g.\ its center lies above or below the mean water level)? As long as the barrier does not create a hole in the fluid domain, transformations such as circle–line or more general Möbius maps can still flatten the geometry to the lower half-plane. In contrast, a fully submerged cylinder yields a multiply connected domain, requiring mappings onto an annulus or strip \citep[e.g.][]{nehari1952conformal, ahlfors1979complex, hassani2013mathematical, caratheodory2021theory}. These more advanced conformal mapping procedures remain theoretically feasible but are mathematically more involved.

(b) \emph{Other body shapes.}
Although linear theory suppresses many geometric details, the circular cylinder case is particularly convenient due to the classical Joukowski transform. For arbitrary shapes, one loses that closed-form map, but the Riemann Mapping Theorem still guarantees the existence of a conformal mapping to the lower half-plane, assuming the fluid region is simply connected. Schwarz–Christoffel transformations \citep{henrici1993applied, driscoll2002schwarz}, successive Möbius mappings \citep{nehari1952conformal}, or numerical schemes \citep{papamichael2010numerical, driscoll2002schwarz} can be employed to solve the linear boundary value problem in a suitably transformed domain.

(c) \emph{Finite water depth.}
Lastly, our model currently assumes infinite (or effectively deep) water depth, which is reasonable in many practical scenarios where wave motion near the free surface is only weakly influenced by a distant seabed. When the depth is not large relative to the wavelength, the finite-depth setting must be considered. Conceptually, one replaces the open half-plane below the free surface with a finite strip bounded by the free surface on top and the bottom boundary beneath. This strip minus the (semi-)cylindrical region remains simply connected, so a suitable mapping still exists, although hyperbolic functions typically arise to satisfy bottom boundary conditions \citep{fetter2003theoretical, mei2005theory}. Extending the present formulation to finite-depth or shallow-water regimes would enable a richer study of capillary–gravity wave scattering across a broader range of physical contexts.

(d) \emph{Meniscus effects.}
All of the above cases assume that the free surface remains flat except for wave-induced deformations, effectively taking a $90^\circ$ static contact angle and thus avoiding a meniscus at the barrier. If this contact angle deviates from $90^\circ$, a static meniscus forms near the barrier, complicating the dynamic boundary condition and altering the fluid domain itself. Even within linear theory, the local curvature induced by the meniscus modifies the boundary condition on the free surface, no longer allowing straightforward conformal mapping to a half-plane minus a semicircle. Analytical expressions for menisci can become cumbersome \citep{landau2013fluid, C-W-book}, and numerical or alternative analytical approaches are often required to treat these geometries accurately. Accounting for meniscus effects in scattering problems represents a significant challenge and a fruitful direction for future study.

(e) \emph{Nonlinear contact line dynamics.}
Real contact line behavior is often governed by hysteresis and nonlinear dependencies between contact line speed and contact angle \citep{ref:Eggers-2005,amberg2022detailed, butt2022contact, eral2013contact}. Such phenomena may entail time-dependent evolutions and energy losses that do not reduce to a simple linear slip parameter. Incorporating these nonlinearities would require a departure from the monochromatic assumption, likely involving unsteady simulations or multi-mode projection methods \citep{bongarzone2021relaxation}. Although challenging, this effort would significantly enhance the realism and predictive capabilities of wave–barrier interaction studies.

Combined with experimental validation and numerical simulations, the research could pave the way for advanced control techniques, improved energy absorption systems, and innovative fluid management strategies in both industrial and research settings. At large scales, our theoretical insights can inform the design of coastal defenses, guiding strategies to reduce wave energy transport and improve stability. At smaller scales, such as in microfluidic devices, manipulating contact line dynamics and barrier geometry can enable precise control of capillary flows and energy partitioning. Understanding and tuning dissipation—interpreted as an impedance-matching condition—can lead to more efficient energy harvesting and fluid control.

\backsection[Funding]{This work was supported by the National Science Foundation (Grant No. 2306106).}

\backsection[Declaration of interests]{The authors report no conflict of interest.}

\backsection[Author ORCIDs]{G. Liu, https://orcid.org/0009-0002-1452-2267; L. zhang, https://orcid.org/0000-0003-3898-6533}

\appendix

\section{Solution to the BVPs for $\eta$}\label{app:solve_eta}

In this appendix, we will solve the inhomogeneous second-order linear differential equation (ISOLDE) for $\eta$ given in \eqref{eq:etaequ} using the method of variation of constants \citep{hassani2013mathematical}. The ISOLDE is given by:
\begin{equation}\label{eq:isolde_app}
    \frac{\mathrm{d}^2 \eta}{\mathrm{d} x^2} - B \eta = - (1+B) \phi(x, 0).
\end{equation}
The corresponding inhomogeneous second-order linear differential equation (HSOLDE), 
\begin{equation}
    \frac{\mathrm{d} ^2 \eta}{\mathrm{d} x^2} - B \eta = 0,
\end{equation}
has two solutions
\begin{equation}
    \eta_1 = \mathrm{e}^{\sqrt{B} x}, \quad \eta_2 = \mathrm{e}^{-\sqrt{B} x}.
\end{equation}
The particular solution to \eqref{eq:isolde_app} can be obtained through variation of constants from $\eta_1, \eta_2$:
\begin{equation}\label{eq:eta*_app}
    \eta^* = \eta_2(x) \int^x \frac{\eta_1(x') r(x')}{W(x')}\:\mathrm{d}x' - \eta_1(x) \int^x \frac{\eta_2(x') r(x')}{W(x')}\:\mathrm{d}x', 
\end{equation}
where $r(x) = - (1+B) \phi(x, 0)$ is the source term in \eqref{eq:isolde_app}, $W(x) = \eta_1 \eta_2' - \eta_1' \eta_2 = - 2 \sqrt{B}$ is the Wronskian of $\eta_1$ and $\eta_2$. Substitute expressions for $\eta_1, \eta_2, r, W$ into \eqref{eq:eta*_app}, we have:
\begin{equation}
    \eta* = \frac{1+B}{2\sqrt{B}} \left( \mathrm{e}^{-\sqrt{B} x} \int^x \phi(x', 0) \mathrm{e}^{\sqrt{B} x'} \:\mathrm{d}x' - \mathrm{e}^{\sqrt{B} x} \int^x \phi(x', 0) \mathrm{e}^{-\sqrt{B} x'} \:\mathrm{d}x' \right).
\end{equation}
Therefore the general solution to \eqref{eq:isolde_app} can be given by:
\begin{equation}
    \eta = c_1 \eta_1 + c_2 \eta_2 + \eta^*, \label{eq:eta_app}
\end{equation}
where $c_1, c_2$ are arbitrary constants. For $x > r$, we have the boundary conditions specified in \eqref{eq:etaequ2}, given by:
\begin{subequations}
    \begin{align}
    \frac{\mathrm{d} \eta}{\mathrm{d} x} &= - \mathrm{i}\Lambda^{-1} \eta, && \text{at } x = r, \\
    \eta &\rightarrow T \mathrm{e}^{ \mathrm{i}x }, && \text{as } x \rightarrow +\infty, 
    \end{align}
\end{subequations}
where the condition at infinity can be further simplified as:
\begin{subequations}\label{eq:bc_app}
    \begin{align}
    \frac{\mathrm{d} \eta}{\mathrm{d} x} &= - \mathrm{i}\Lambda^{-1} \eta, && \text{at } x = r, \\
    \frac{\mathrm{d} \eta}{\mathrm{d} x} &= \mathrm{i} \eta, && \text{as } x \rightarrow +\infty. 
    \end{align}
\end{subequations}
From \eqref{eq:eta_app}, we have the form for $\eta$ and $\mathrm{d} \eta / \mathrm{d} x$:
\begin{subequations}\label{eq:eta&eta'_app}
\begin{align}\label{eq:eta+_app}
    \eta &= c_1 \mathrm{e}^{\sqrt{B} x} + c_2 \mathrm{e}^{-\sqrt{B} x} \nonumber \\
    &\quad + \frac{1+B}{2\sqrt{B}} \left( \mathrm{e}^{-\sqrt{B} x} \int_r^x \phi(x', 0) \mathrm{e}^{\sqrt{B} x'} \:\mathrm{d}x' - \mathrm{e}^{\sqrt{B} x} \int_r^x \phi(x', 0) \mathrm{e}^{-\sqrt{B} x'} \:\mathrm{d}x' \right),
\end{align}
and 
\begin{align}
    \frac{\mathrm{d} \eta}{\mathrm{d} x} &= \sqrt{B} c_1 \mathrm{e}^{\sqrt{B} x} - \sqrt{B} c_2 \mathrm{e}^{-\sqrt{B} x} \nonumber \\
    &\quad - \frac{1+B}{2} \left( \mathrm{e}^{-\sqrt{B} x} \int_r^x \phi(x', 0) \mathrm{e}^{\sqrt{B} x'} \:\mathrm{d}x' + \mathrm{e}^{\sqrt{B} x} \int_r^x \phi(x', 0) \mathrm{e}^{-\sqrt{B} x'} \:\mathrm{d}x' \right).
\end{align}
\end{subequations}
Substitute \eqref{eq:eta&eta'_app} to \eqref{eq:bc_app}, we get:
\begin{subequations}
    \begin{align}
        c_1 &= \frac{1+B}{2\sqrt{B}} \int_r^\infty \phi(x', 0) \mathrm{e}^{-\sqrt{B} x'} \:\mathrm{d}x', \\
        c_2 &= \frac{\sqrt{B} \Lambda + \mathrm{i}}{\sqrt{B} \Lambda - \mathrm{i}} \mathrm{e}^{2 \sqrt{B} r} c_1.
    \end{align}
\end{subequations}
Therefore, if we define the slip parameter:
\begin{equation}
    \zeta = \tan^{-1} \left(\sqrt{B} \Lambda\right),
\end{equation}
$c_2$ becomes:
\begin{align}
    c_2 &= \frac{\tan\zeta + \mathrm{i}}{\tan\zeta - \mathrm{i}} \mathrm{e}^{2 \sqrt{B} r} c_1 \nonumber\\
    &= \mathrm{e}^{2\mathrm{i} (\frac{\pi}{2} - \zeta)} \mathrm{e}^{2 \sqrt{B} r} c_1 \nonumber\\
    &= - \frac{1+B}{2\sqrt{B}} \mathrm{e}^{-2\mathrm{i} \zeta} \mathrm{e}^{2 \sqrt{B} r} \int_r^\infty \phi(x', 0) \mathrm{e}^{-\sqrt{B} x'} \:\mathrm{d}x'.
\end{align}
Substitute $c_1, c_2$ to \eqref{eq:eta+_app}:
\begin{align}
    \eta &= \frac{1+B}{2\sqrt{B}} \bigg( \mathrm{e}^{\sqrt{B} x} \int_r^\infty \phi(x', 0) \mathrm{e}^{-\sqrt{B} x'} \:\mathrm{d}x' - \mathrm{e}^{-2\mathrm{i} \zeta} \mathrm{e}^{2 \sqrt{B} (2r-x)} \int_r^\infty \phi(x', 0) \mathrm{e}^{-\sqrt{B} x'} \:\mathrm{d}x' \nonumber \\
    &\qquad\quad\  + \mathrm{e}^{-\sqrt{B} x} \int_r^x \phi(x', 0) \mathrm{e}^{\sqrt{B} x'} \:\mathrm{d}x' - \mathrm{e}^{\sqrt{B} x} \int_r^x \phi(x', 0) \mathrm{e}^{-\sqrt{B} x'} \:\mathrm{d}x' \bigg) \nonumber \\
    &= \frac{1+B}{2\sqrt{B}} \bigg( \int_r^x \phi(x', 0) \mathrm{e}^{\sqrt{B} (x'-x)} \:\mathrm{d}x' + \int_x^\infty \phi(x', 0) \mathrm{e}^{\sqrt{B} (x-x')} \:\mathrm{d}x' \nonumber \\
    &\qquad\quad\  - \mathrm{e}^{-2\mathrm{i} \zeta} \int_r^\infty \phi(x', 0) \mathrm{e}^{\sqrt{B} (2r-(x+x'))} \:\mathrm{d}x' \bigg) \nonumber \\
    &= \int_r^\infty \frac{1+B}{2\sqrt{B}} \left( - \mathrm{e}^{-2\mathrm{i} \zeta} \mathrm{e}^{\sqrt{B} (2r-(x+x'))} +\mathrm{e}^{-\sqrt{B} |x-x'|}  \right) \phi(x', 0) \:\mathrm{d}x'.\label{eq:etasol+_app}
\end{align}

For $x < -r$, following the same approach, we have:
\begin{equation}
    \eta = \int_{-\infty}^{-r} \frac{1+B}{2\sqrt{B}}  \left( - \mathrm{e}^{-2\mathrm{i} \zeta} \mathrm{e}^{\sqrt{B} (2r+(x+x'))} +\mathrm{e}^{-\sqrt{B} |x-x'|}  \right) \phi(x', 0) \:\mathrm{d}x'.\label{eq:etasol-_app}
\end{equation}
Combine \eqref{eq:etasol+_app} and \eqref{eq:etasol-_app}, we have the solution for $\eta$ in general:
\begin{equation}
\eta = 
\begin{cases}
\displaystyle \int^{- r}_{- \infty} K(x, x') \phi(x', 0) \: \mathrm{d} x', & \forall x < -r,  \\
\displaystyle \int^{+ \infty}_{r} K(x, x') \phi(x', 0) \: \mathrm{d} x', & \forall x > r,
\end{cases}
\end{equation}
where the dynamic kernel $K(x, x')$ is defined as:
\begin{equation}
K(x, x') = \frac{1+B}{2\sqrt{B}} \left( -\mathrm{e}^{-2\mathrm{i}\zeta} \mathrm{e}^{\sqrt{B}(2r - |x + x'|)} + \mathrm{e}^{-\sqrt{B}|x - x'|} \right).
\end{equation}

\section{Mathematical Details in Expressing $\phi(x,0)$}\label{app:solve_phi}

This appendix provides the detailed steps required to derive $\phi(x,0)$ in terms of $\eta(x)$, transitioning from \eqref{eq:phi0sol_temp} to \eqref{eq:phi0sol}. 

Recall \eqref{eq:phi0sol_temp}:
\begin{equation}\label{eq:phi0sol_temp_app}
\psi(u, v) 
= \frac{1}{2\upi} \int_{-\infty}^{+\infty} g(u') \biggl(\,\int_{-\infty}^{+\infty} \frac{1}{|\lambda|} \mathrm{e}^{|\lambda| v} \,\mathrm{e}^{\mathrm{i}\lambda (u - u')} \,\mathrm{d}\lambda \biggr) \mathrm{d}u'.
\end{equation}
Evaluating at $v=0$ gives
\begin{align}\label{eq:phi0sol_app1}
\psi(u, 0) 
&= \int_{-\infty}^{+\infty} g(u') \biggl( \frac{1}{2\upi} \int_{-\infty}^{+\infty} \frac{1}{|\lambda|} \mathrm{e}^{\mathrm{i}\lambda (u - u')} \,\mathrm{d}\lambda \biggr) \mathrm{d}u'.
\end{align}
Recognize that the expression in parentheses is the inverse Fourier transform of $1/|\lambda|$, namely
\begin{equation}
    \frac{1}{2\upi} \int_{-\infty}^{+\infty} \frac{1}{|\lambda|} \,\mathrm{e}^{\mathrm{i}\lambda (u - u')} \,\mathrm{d}\lambda 
    \;=\; \mathcal{F}^{-1}\!\Bigl[\frac{1}{|\lambda|}\Bigr](u - u')
    \;=\; -\,\frac{1}{\upi}\,\bigl(\ln|u - u'|\,+\,\gamma\bigr),
\end{equation}
where $\gamma$ is the Euler--Mascheroni constant. Substituting this back into \eqref{eq:phi0sol_app1} gives
\begin{equation}
    \psi(u, 0) 
    \;=\;
    -\,\frac{1}{\upi}
    \int_{-\infty}^{+\infty} 
    g(u')
    \bigl(\ln|u - u'| + \gamma\bigr)\,\mathrm{d}u'.
\end{equation}

Since $\psi(u, v) = \phi(x, y)$ under the conformal mapping, we obtain
\begin{equation}
    \phi(x, 0) 
    \;=\; \psi\bigl(u(x,0),\,v(x,0)\bigr)
    \;=\;
    \psi\bigl(u(x,0),\,0\bigr).
\end{equation}
From the conformal mapping in component form in \eqref{eq:conformal_components}, we have
\begin{equation}
u = \frac{(x^2 + y^2 + r^2)\,x}{(x^2 + y^2)\,r},
\quad
v = \frac{(x^2 + y^2 - r^2)\,y}{(x^2 + y^2)\,r}.
\end{equation}
Thus, for $y=0$,
\begin{equation}
    u(x,0) \;=\; \frac{(x^2 + r^2)\,x}{x^2\,r}
    \;=\; \frac{x}{r} + \frac{r}{x}.
\end{equation}

Next, recall the relation between the boundary function $g(u)$ and the free-surface elevation $\eta(x)$ from \eqref{eq:g2eta}:
\begin{equation}
g(u) = 
\begin{cases}
\displaystyle 0, & \mbox{for } |u| < 2,  \\
\displaystyle \eta / \partial_x u(x, 0), & \mbox{for } |u| > 2,
\end{cases}
\end{equation}
which implies
\begin{equation}
    g(u)\:\mathrm{d}u 
    \;=\; \eta(x)\:\mathrm{d}x,
    \quad
    \text{for } |u| > 2 \;\;\text{(equivalently, } |x| > r\text{).}
\end{equation}
Hence we obtain
\begin{align}
    \phi(x, 0) &= \psi(u(x,0), 0) \nonumber \\
    &= - \upi^{-1} \left( \int_{-\infty}^{-2} + \int_{r}^{2} \right) \, \left( \ln{|u(x,0) - u'(x',0)|} + \gamma \right) g(u') \: \mathrm{d} u' \nonumber \\
    &= - \upi^{-1} \left( \int_{-\infty}^{-r} + \int_{r}^{+\infty} \right) \, \left( \ln{ \left| \frac{x}{r} + \frac{r}{x} - \frac{x'}{r} - \frac{r}{x'} \right| } + \gamma \right) \eta(x') \: \mathrm{d} x' .
\end{align}

Therefore, for $|x|>r$, we arrive at the final expression \eqref{eq:phi0sol}:
\begin{equation}
\phi(x, 0)
\;=\;
\left( \int_{-\infty}^{-r} + \int_{r}^{+\infty} \right)
S(x, x')\,\eta(x')\,\mathrm{d}x',
\end{equation}
where the geometric kernel $S(x, x')$ is given by
\begin{equation}
S (x, x') = - \upi^{-1} \left( \ln{ \left| \frac{x}{r} + \frac{r}{x} - \frac{x'}{r} - \frac{r}{x'} \right| } + \gamma \right).
\end{equation}
Hence, $\phi(x,0)$ is expressed solely in terms of $\eta(x)$ through the integral kernel $S(x,x')$.

\section{Analytical Calculation of $(\mathcal{L}_{\chi} \mathcal{K} - \mathcal{I}) \mathrm{e}^{\pm \mathrm{i}x}$}\label{app:expression}

In this appendix, we will analytically calculate $(\mathcal{L}_{\chi} \mathcal{K} - \mathcal{I}) \mathrm{e}^{\pm \mathrm{i}x}$ that proposed in \eqref{eq:(LK-I)exp}, where $\chi$ takes the values `$e$' for even and `$o$' for odd, $\mathcal{L}_e$, $\mathcal{L}_o$ and $\mathcal{K}$ are the integral operators integrate from $r$ to $\infty$ with kernels $L_e$, $L_o$ and $K$ given by:
\begin{subequations}
    \begin{align}
L_e(x,x') 
&= S(x, x') + S(x, -x') , \\
L_o(x,x') 
&= S(x, x') - S(x, -x') , \\
K(x,x')
&= \frac{1+B}{2\sqrt{B}} \left( -\mathrm{e}^{-2\mathrm{i}\zeta} \mathrm{e}^{\sqrt{B}(2r - |x + x'|)} + \mathrm{e}^{-\sqrt{B}|x - x'|} \right) ,
\end{align}
where
\begin{equation}
    S(x, x') = - \upi^{-1} \left( \ln \left| \frac{x}{r} + \frac{r}{x} - \frac{x'}{r} - \frac{r}{x'} \right| + \gamma \right) .
\end{equation}
\end{subequations}

The expression $(\mathcal{L}_{\chi} \mathcal{K} - \mathcal{I}) \mathrm{e}^{\pm \mathrm{i}x}$ can be written in integral form:
\begin{align}\label{eq:(LK-I)exp_appA}
    (\mathcal{L}_{\chi} \mathcal{K} - \mathcal{I}) \mathrm{e}^{\pm \mathrm{i}x} &= \int_{r}^{+ \infty} L_{\chi}(x, x') \left( \int_{r}^{+ \infty} K(x', x'') \mathrm{e}^{\pm \mathrm{i}x''} \: \mathrm{d} x'' \right) \mathrm{d} x' - \mathrm{e}^{\pm \mathrm{i}x} .
\end{align}

First, we calculate $\int_{r}^{+ \infty} K(x', x'') \mathrm{e}^{\pm \mathrm{i}x''} \mathrm{d} x''$:
\begin{align}
    \int_{r}^{+ \infty} K(x', x'') \mathrm{e}^{\pm \mathrm{i}x''} \mathrm{d} x'' 
    &= \int_{r}^{+ \infty} \frac{1+B}{2\sqrt{B}} \left( -\mathrm{e}^{-2\mathrm{i}\zeta} \mathrm{e}^{\sqrt{B}(2r - |x' + x''|)} + \mathrm{e}^{-\sqrt{B}|x' - x''|} \right) \mathrm{e}^{\pm \mathrm{i}x''} \mathrm{d} x''  \nonumber \\
    &= \frac{1}{2 \sqrt{B}} \left[ - (\sqrt{B} \pm \mathrm{i}) \mathrm{e}^{-2\mathrm{i}\zeta} + (-\sqrt{B} \pm \mathrm{i}) \right] \mathrm{e}^{(\sqrt{B} \pm \mathrm{i}) r} \mathrm{e}^{- \sqrt{B} x'} + \mathrm{e}^{\pm \mathrm{i}x'} \label{eq:J2} .
\end{align}

Substitute \eqref{eq:J2} into \eqref{eq:(LK-I)exp_appA}, we have:
\begin{equation}\label{eq:(LK-I)exp2_app}
    (\mathcal{L}_{\chi} \mathcal{K} - \mathcal{I}) \mathrm{e}^{\pm \mathrm{i}x}
    = - \mathrm{e}^{\pm \mathrm{i}x} + \frac{1}{2 \sqrt{B}} \left[ - (\sqrt{B} \pm \mathrm{i}) \mathrm{e}^{-2\mathrm{i}\zeta} + (-\sqrt{B} \pm \mathrm{i}) \right] \mathrm{e}^{(\sqrt{B} \pm \mathrm{i}) r} I_{1, \chi} + I_{2, \chi, \pm} ,
\end{equation}
where
\begin{subequations}
    \begin{align}
    I_{1, \chi} &= \int_{r}^{+ \infty} L_{\chi}(x, x') \mathrm{e}^{- \sqrt{B} x'} \mathrm{d} x' , \\
    I_{2, \chi, \pm} &= \int_{r}^{+ \infty} L_{\chi}(x, x') \mathrm{e}^{\pm \mathrm{i}x'} \mathrm{d} x' .
\end{align}
\end{subequations}

Since $L_e = S(x, x') + S(x, -x')$ and $L_e = S(x, x') - S(x, -x')$, first we calculate:
\begin{align}
    \int_{r}^{+ \infty} S(x, x') \mathrm{e}^{- \sqrt{B} x'} \mathrm{d} x' &= - \upi^{-1} \int_{r}^{+ \infty} \left( \ln{ \left| \frac{x}{r} + \frac{r}{x} - \frac{x'}{r} - \frac{r}{x'} \right| } + \gamma \right) \mathrm{e}^{- \sqrt{B} x'} \mathrm{d} x'  \nonumber \\
    &= \frac{1}{\upi \sqrt{B}} \bigg( \mathrm{e}^{-\frac{\sqrt{B} r^2}{x}} \operatorname{Ei}\left( -\frac{\sqrt{B} r (x-r)}{x} \right) + \mathrm{e}^{-\sqrt{B} x} \operatorname{Ei}(\sqrt{B}(x-r)) \nonumber \\
    &\qquad\quad\  -\operatorname{Ei}(-\sqrt{B} r) - \ln{ \left| \frac{x}{r} + \frac{r}{x} - 2 \right| } \mathrm{e}^{-\sqrt{B} r} - \gamma \mathrm{e}^{-\sqrt{B} r} \bigg) . 
\end{align}
Since $S(x, -x') = S(-x, x')$, we have:
\begin{align}
    \int_{r}^{+ \infty} S(x, -x') \mathrm{e}^{- \sqrt{B} x'} \mathrm{d} x' &= \int_{r}^{+ \infty} S(-x, x') \mathrm{e}^{- \sqrt{B} x'} \mathrm{d} x'  \nonumber \\
    &=\frac{1}{\upi \sqrt{B}} \bigg( \mathrm{e}^{\frac{\sqrt{B} r^2}{x}} \operatorname{Ei}\left( -\frac{\sqrt{B} r (x+r)}{x} \right) + \mathrm{e}^{\sqrt{B} x} \operatorname{Ei}(-\sqrt{B}(x+r)) \nonumber \\
    &\qquad\quad\  -\operatorname{Ei}(-\sqrt{B} r) - \ln{ \left| \frac{x}{r} + \frac{r}{x} + 2 \right| } \mathrm{e}^{-\sqrt{B} r} - \gamma \mathrm{e}^{-\sqrt{B} r} \bigg) . 
\end{align}
Now we calculate $I_1$:
\begin{subequations}\label{eq:I1_app}
\begin{align}
    I_{1, e} &= \int_{r}^{+ \infty} L_{e}(x, x') \mathrm{e}^{- \sqrt{B} x'} \mathrm{d} x'  \nonumber \\
    &= \int_{r}^{+ \infty} S(x, x') \mathrm{e}^{- \sqrt{B} x'} \mathrm{d} x' + \int_{r}^{+ \infty} S(x, -x') \mathrm{e}^{- \sqrt{B} x'} \mathrm{d} x'  \nonumber \\
    &= \frac{1}{\upi \sqrt{B}} \bigg( \mathrm{e}^{-\frac{\sqrt{B} r^2}{x}} \operatorname{Ei}\left( -\frac{\sqrt{B} r (x-r)}{x} \right) + \mathrm{e}^{\frac{\sqrt{B} r^2}{x}} \operatorname{Ei}\left( -\frac{\sqrt{B} r (x+r)}{x} \right) \nonumber \\
    &\qquad\qquad + \mathrm{e}^{-\sqrt{B} x} \operatorname{Ei}(\sqrt{B}(x-r)) + \mathrm{e}^{\sqrt{B} x} \operatorname{Ei}(-\sqrt{B}(x+r)) \nonumber \\
    &\qquad\qquad - 2 \operatorname{Ei}(-\sqrt{B} r) - 2 \ln{ \left( \frac{x}{r} - \frac{r}{x} \right) } \mathrm{e}^{-\sqrt{B} r} - 2 \gamma \mathrm{e}^{-\sqrt{B} r} \bigg) , \\
    I_{1, o} &= \int_{r}^{+ \infty} L_{o}(x, x') \mathrm{e}^{- \sqrt{B} x'} \mathrm{d} x'  \nonumber \\
    &= \int_{r}^{+ \infty} S(x, x') \mathrm{e}^{- \sqrt{B} x'} \mathrm{d} x' - \int_{r}^{+ \infty} S(x, -x') \mathrm{e}^{- \sqrt{B} x'} \mathrm{d} x'  \nonumber \\
    &= \frac{1}{\upi \sqrt{B}} \bigg( \mathrm{e}^{-\frac{\sqrt{B} r^2}{x}} \operatorname{Ei}\left( -\frac{\sqrt{B} r (x-r)}{x} \right) - \mathrm{e}^{\frac{\sqrt{B} r^2}{x}} \operatorname{Ei}\left( -\frac{\sqrt{B} r (x+r)}{x} \right) \nonumber \\
    &\qquad\qquad + \mathrm{e}^{-\sqrt{B} x} \operatorname{Ei}(\sqrt{B}(x-r)) - \mathrm{e}^{\sqrt{B} x} \operatorname{Ei}(-\sqrt{B}(x+r)) \nonumber \\
    &\qquad\qquad - 2 \ln{ \left( \frac{x - r}{x + r} \right) } \mathrm{e}^{-\sqrt{B} r} \bigg) .
\end{align}
\end{subequations}

Some care is needed in the calculation of $I_2$. Since $\mathrm{e}^{\pm \mathrm{i}x}$ is $O(1)$ and $L_\chi$ is $O(\ln x)$ at $x \rightarrow \infty$, $I_2$ is not converged on range $[r, \infty]$. Some mathematical technique is used to calculate the principal value of $I_2$, as shown below:
\begin{subequations}\label{eq:I2_app}
\begin{align}
    I_{2, e, \pm} &= P \int_{r}^{+ \infty} L_{e}(x, x') \mathrm{e}^{\pm \mathrm{i}x'} \mathrm{d} x'  \nonumber \\
    &= \lim_{\epsilon \rightarrow 0^+} \int_{r}^{+ \infty} L_{e}(x, x') \mathrm{e}^{\pm \mathrm{i}x'} \mathrm{e}^{- \epsilon x'} \mathrm{d} x'  \nonumber \\
    &=\frac{1}{\upi} \Bigg( \mp 2\mathrm{i}\gamma \mathrm{e}^{\pm \mathrm{i}r} - 2\upi + \upi \mathrm{e}^{\mp \mathrm{i}x} + 2\upi \cos\left(\frac{r^2}{x}\right) \mp 2\mathrm{i}\operatorname{Ei}(\pm \mathrm{i}r) \nonumber \\
&\quad \pm \mathrm{i}\bigg[ \mathrm{e}^{\frac{\mathrm{i}r^2}{x}} \operatorname{Ei}\left(\frac{\mathrm{i}r(\pm x-r)}{x}\right) + \mathrm{e}^{-\frac{\mathrm{i}r^2}{x}} \operatorname{Ei}\left(\frac{\mathrm{i}r(\pm x+r)}{x}\right) \nonumber \\
&\qquad \ + \mathrm{e}^{\mathrm{i}x} \operatorname{Ei}(\mathrm{i}(-x\pm r)) + \mathrm{e}^{-\mathrm{i}x} \operatorname{Ei}(\mathrm{i}(x\pm r)) \nonumber \\
&\qquad \ - 2\mathrm{e}^{\pm \mathrm{i}r} \ln\left(\frac{x}{r}-\frac{r}{x}\right)\bigg] \Bigg) , \\
I_{2, o, \pm} &= P \int_{r}^{+ \infty} L_{o}(x, x') \mathrm{e}^{\pm \mathrm{i}x'} \mathrm{d} x' \nonumber \\
    &= \lim_{\epsilon \rightarrow 0^+} \int_{r}^{+ \infty} L_{o}(x, x') \mathrm{e}^{\pm \mathrm{i}x'} \mathrm{e}^{- \epsilon x'} \mathrm{d} x' \nonumber \\
    &= \frac{1}{\upi} \Bigg( - \upi \mathrm{e}^{\mp \mathrm{i}x} \pm 2\upi \mathrm{i}\sin\left(\frac{r^2}{x}\right) \nonumber \\
&\quad + \mathrm{i}\bigg[ \mathrm{e}^{\frac{\mathrm{i}r^2}{x}} \operatorname{Ei}\left(\frac{\mathrm{i}r(\pm x-r)}{x}\right) - \mathrm{e}^{-\frac{\mathrm{i}r^2}{x}} \operatorname{Ei}\left(\frac{\mathrm{i}r(\pm x+r)}{x}\right) \nonumber \\
&\qquad \ + \mathrm{e}^{\mathrm{i}x} \operatorname{Ei}(\mathrm{i}(-x\pm r)) - \mathrm{e}^{- \mathrm{i}x} \operatorname{Ei}(\mathrm{i}(x\pm r)) \nonumber \\
&\qquad \ \mp 2\mathrm{e}^{\pm \mathrm{i}r} \ln\left(\frac{x-r}{x+r}\right)\bigg] \Bigg) .
\end{align}
\end{subequations}

Now we have the analytical expression for $(\mathcal{L}_{\chi} \mathcal{K} - \mathcal{I}) \mathrm{e}^{\pm \mathrm{i}x}$, recalling \eqref{eq:(LK-I)exp2_app}:
\begin{equation}
    (\mathcal{L}_{\chi} \mathcal{K} - \mathcal{I}) \mathrm{e}^{\pm \mathrm{i}x}
    = - \mathrm{e}^{\pm \mathrm{i}x} + \frac{1}{2 \sqrt{B}} \left[ - (\sqrt{B} \pm \mathrm{i}) \mathrm{e}^{-2\mathrm{i}\zeta} + (-\sqrt{B} \pm \mathrm{i}) \right] \mathrm{e}^{(\sqrt{B} \pm \mathrm{i}) r} I_{1, \chi} + I_{2, \chi, \pm} ,
\end{equation}
with $I_1, I_2$ defined in \eqref{eq:I1_app} and \eqref{eq:I2_app}.

\section{Analytical Solution for $T$, $R$ in Scattering by an Infinitesimal Barrier}\label{app:solution}

In this appendix, we will solve integral equations proposed in \eqref{eq:heopmequ_infi}:
\begin{equation}\label{eq:heopmequ_appB}
h_{\chi, \pm} = \mathcal{L}_{\chi} \mathcal{K} h_{\chi,\pm} + g_{\chi, \pm} ,
\end{equation}
where $\chi$ takes the values `$e$' for even and `$o$' for odd, $\mathcal{L}_e$, $\mathcal{L}_o$ and $\mathcal{K}$ are the integral operators integrate from $0$ to $\infty$ with kernels $L_e$, $L_o$ and $K$ given by:
\begin{subequations}
\begin{align}
L_e(x,x') 
&= - \upi^{-1} \left( \ln \left| x - x' \right| + \ln \left( x + x' \right) + 2 \gamma \right) , \\
L_o(x,x') 
&= - \upi^{-1} \left( \ln \left| x - x' \right| - \ln \left( x + x' \right) \right) , \\
K(x,x')
&= \frac{1+B}{2\sqrt{B}} \left( -\mathrm{e}^{-2\mathrm{i}\zeta} \mathrm{e}^{- \sqrt{B} (x + x')} + \mathrm{e}^{-\sqrt{B}|x - x'|} \right) ,
\end{align}
\end{subequations}
and $g_{\chi, \pm} = (\mathcal{L}_{\chi} \mathcal{K} - \mathcal{I}) \mathrm{e}^{\pm \mathrm{i}x}$ can be calculated analytically through similar methods as in Appendix~\ref{app:expression}:
\begin{align}
g_{e, \pm} &= (\mathcal{L}_{e} \mathcal{K} - \mathcal{I}) \mathrm{e}^{\pm \mathrm{i}x} \nonumber \\
&=\int_{0}^{+ \infty} L_{e}(x, x') \left( \int_{0}^{+ \infty} K(x', x'') \mathrm{e}^{\pm \mathrm{i}x''} \mathrm{d} x'' \right) \mathrm{d} x' - \mathrm{e}^{\pm \mathrm{i}x} \nonumber \\
&= \mp 2 \mathrm{i}\sin x \pm \frac{\mathrm{i}}{\upi} \left[ \mathrm{e}^{\mathrm{i}x} \operatorname{Ei}(- \mathrm{i}x) + \mathrm{e}^{- \mathrm{i}x} \operatorname{Ei}(\mathrm{i}x) - 2 (\gamma + \ln |x|) \right] \nonumber \\
&\quad + \frac{- (\sqrt{B} \pm \mathrm{i}) \mathrm{e}^{- 2 \mathrm{i}\zeta} + (- \sqrt{B} \pm \mathrm{i})}{2 \upi B} \left[ \mathrm{e}^{\sqrt{B} x} \operatorname{Ei}(- \sqrt{B} x) + \mathrm{e}^{- \sqrt{B} x} \operatorname{Ei}(\sqrt{B} x) - 2 (\gamma + \ln |x|) \right], \\
g_{o, \pm} &= (\mathcal{L}_{o} \mathcal{K} - \mathcal{I}) \mathrm{e}^{\pm \mathrm{i}x} \nonumber \\
&=\int_{0}^{+ \infty} L_{o}(x, x') \left( \int_{0}^{+ \infty} K(x', x'') \mathrm{e}^{\pm \mathrm{i}x''} \mathrm{d} x'' \right) \mathrm{d} x' - \mathrm{e}^{\pm \mathrm{i}x} \nonumber \\
&= - 2 \cos x + \frac{\mathrm{i}}{\upi} \left[ \mathrm{e}^{\mathrm{i}x} \operatorname{Ei}(- \mathrm{i}x) - \mathrm{e}^{- \mathrm{i}x} \operatorname{Ei}(\mathrm{i}x) \right] \nonumber \\
&\quad + \frac{- (\sqrt{B} \pm \mathrm{i}) \mathrm{e}^{- 2 \mathrm{i}\zeta} + (- \sqrt{B} \pm \mathrm{i})}{2 \upi B} \left[ - \mathrm{e}^{\sqrt{B} x} \operatorname{Ei}(- \sqrt{B} x) + \mathrm{e}^{- \sqrt{B} x} \operatorname{Ei}(\sqrt{B} x) \right].
\end{align}

First, we solve for $h_e$. We apply a Fourier Cosine Transform (FCT) 
\begin{equation}
    \hat{f}_c(k) = \mathcal{F}_c\left\{ f \right\} = \int_0^\infty f(x) \cos kx \: \mathrm{d} x ,
\end{equation}
to \eqref{eq:heopmequ_appB}, we obtain:
\begin{equation}\label{eq:hcequ}
    \hat{h}_{e,c}(k) = \hat{g}_{e,c}(k) + \mathcal{F}_c\left\{\int_0^\infty L_e(x, x') \mathcal{K} h_e(x') dx' \right\} .
\end{equation}
The second term on the right side can be expanded as:
\begin{align}
&\quad \mathcal{F}_c\left\{\int_0^\infty L_e(x, x')\mathcal{K} h_e(x') \: \mathrm{d} x' \right\} \nonumber \\
&= \mathcal{F}_c\left\{ \int_0^\infty - \upi^{-1} \left( \ln \left| x - x' \right| + \ln \left( x + x' \right) + 2 \gamma \right) \mathcal{K} h_e(x') \: \mathrm{d} x' \right\} \nonumber \\
&= - \upi^{-1} \mathcal{F}_c\left\{ \int_0^\infty \mathcal{K} h_e(x') \left[ ( \ln \left| x - x' \right| + \gamma ) + ( \ln \left( x + x' \right) + \gamma ) \right] \: \mathrm{d} x'\right\} . 
\end{align}
For FCT, we have the convolution theorem:
\begin{equation}
    2 \mathcal{F}_c\left\{ f \right\} \mathcal{F}_c\left\{ g \right\} = \mathcal{F}_c\left\{ \int_0^\infty f(x') \left[ g(x + x') + g(|x - x'|) \right] \: \mathrm{d} x' \right\}.
\end{equation}
Therefore
\begin{align}
&\quad \mathcal{F}_c\left\{\int_0^\infty L_e(x, x') \mathcal{K} h_e(x') \: \mathrm{d} x' \right\} \nonumber \\
&= - \frac{2}{\upi} \mathcal{F}_c\left\{ \ln x + \gamma \right\} \mathcal{F}_c\left\{ \mathcal{K} h_e \right\} \nonumber \\
&= - \frac{2}{\upi} \left( - \frac{\upi}{2k} \mathcal{F}_c\left\{ \mathcal{K} h_e \right\} \right) \nonumber \\
&= \frac{1}{k} \mathcal{F}_c\left\{ \mathcal{K} h_e \right\} .
\end{align}
The FCT of $\mathcal{K} h_e$ can be calculated by
\begin{align}
\mathcal{F}_c\left\{ \mathcal{K} h_e \right\} &= \mathcal{F}_c\left\{ \int_0^\infty \frac{1+B}{2\sqrt{B}} \left( -\mathrm{e}^{-2\mathrm{i}\zeta} \mathrm{e}^{- \sqrt{B} (x + x')} + \mathrm{e}^{-\sqrt{B}|x - x'|} \right) h_e(x') \: \mathrm{d} x' \right\} \nonumber \\
&= \frac{1+B}{2\sqrt{B}} \bigg( \mathcal{F}_c\left\{ \int_0^\infty \left( \mathrm{e}^{- \sqrt{B} (x + x')} + \mathrm{e}^{-\sqrt{B}|x - x'|} \right) h_e(x') \: \mathrm{d} x' \right\} \nonumber \\
&\qquad \qquad - (1 + \mathrm{e}^{-2 \mathrm{i}\zeta}) \mathcal{F}_c\left\{ \int_0^\infty \mathrm{e}^{- \sqrt{B} (x + x')} h_e(x') \: \mathrm{d} x' \right\} \bigg) \nonumber \\
&= \frac{1+B}{2\sqrt{B}} \left( 2 \mathcal{F}_c\left\{ \mathrm{e}^{- \sqrt{B} x} \right\} \hat{h}_{e,c}(k) - (1 + \mathrm{e}^{-2 \mathrm{i}\zeta}) \mathcal{F}_c\left\{ \mathrm{e}^{- \sqrt{B} x} \right\} \int_0^\infty \mathrm{e}^{- \sqrt{B} x'} h_e(x') \: \mathrm{d} x' \right) \nonumber \\
&= \frac{1+B}{2\sqrt{B}} \left( 2 \frac{\sqrt{B}}{B + k^2} \hat{h}_{e,c}(k) - (1 + \mathrm{e}^{-2 \mathrm{i}\zeta}) \frac{\sqrt{B}}{B + k^2} \int_0^\infty \mathrm{e}^{- \sqrt{B} x'} h_e(x') \: \mathrm{d} x' \right) \nonumber \\
&= \frac{1+B}{B+k^2} \left( \hat{h}_{e,c}(k) - \frac{1 + \mathrm{e}^{-2 \mathrm{i}\zeta}}{2} \int_0^\infty \mathrm{e}^{- \sqrt{B} x'} h_e(x') \: \mathrm{d} x' \right) .
\end{align}
Substitute all these back into \eqref{eq:hcequ}, we have:
\begin{equation}\label{eq:hcequ2}
\hat{h}_{e,c}(k) = \hat{g}_{e,c}(k) + \frac{1+B}{k(B+k^2)} \left( \hat{h}_{e,c}(k) - \frac{1 + \mathrm{e}^{-2 \mathrm{i}\zeta}}{2} \int_0^\infty \mathrm{e}^{- \sqrt{B} x'} h_e(x') \: \mathrm{d} x' \right) .
\end{equation}
We want to express the integral on the right side in $k$-domain, we can write $h$ with an inverse FCT:
\begin{equation}
    h_e(x) = \frac{2}{\upi} \int_0^\infty \hat{h}_{e,c}(k) \cos kx \: \mathrm{d} k,
\end{equation}
therefore the integral in \eqref{eq:hcequ2} can be written as:
\begin{align}
\int_0^\infty \mathrm{e}^{- \sqrt{B} x'} h_e(x') \: \mathrm{d} x' &= \int_0^\infty \mathrm{e}^{- \sqrt{B} x'} \left( \frac{2}{\upi} \int_0^\infty \hat{h}_{e,c}(k) \cos kx' \: \mathrm{d} k \right) \mathrm{d} x' \nonumber \\
&= \frac{2}{\upi} \int_0^\infty \hat{h}_{e,c}(k) \left( \int_0^\infty \mathrm{e}^{- \sqrt{B} x'} \cos kx' \: \mathrm{d} x' \right) \mathrm{d} k \nonumber \\
&= \frac{2}{\upi} \int_0^\infty \hat{h}_{e,c}(k) \mathcal{F}_c\left\{ \mathrm{e}^{- \sqrt{B} x} \right\} \: \mathrm{d} k \nonumber \\
&= \frac{2}{\upi} \int_0^\infty \frac{\sqrt{B}}{B+k^2} \hat{h}_{e,c}(k) \: \mathrm{d} k .
\end{align}
Substitute back to \eqref{eq:hcequ2}, we obtain
\begin{equation}
    \hat{h}_{e,c}(k) = \hat{g}_{e,c}(k) + \frac{1+B}{k(B+k^2)} \left( \hat{h}_{e,c}(k) - \frac{1 + \mathrm{e}^{-2 \mathrm{i}\zeta}}{\upi} \int_0^\infty \frac{\sqrt{B}}{B+k^2} \hat{h}_{e,c}(k) \: \mathrm{d} k \right) ,
\end{equation}
which can be rearranged as:
\begin{subequations}
\begin{equation}\label{eq:hcequ3}
    \hat{h}_{e,c}(k) = c \delta(k-1) + k (B+k^2) \hat{w}_c(k) \hat{g}_{e,c}(k) - \alpha \hat{w}_c(k) \int_0^\infty \frac{\hat{h}_{e,c}(k)}{B+k^2} \: \mathrm{d} k ,
\end{equation}
where $c$ is a constant to be determined, and
\begin{align}
\alpha &= \upi^{-1} \sqrt{B} (1+B) (1 + \mathrm{e}^{-2 \mathrm{i}\zeta}) , \\
\hat{w}_c(k) &= (k-1)^{-1}(k^2 + k + 1 + B)^{-1} .
\end{align}
\end{subequations}
\eqref{eq:hcequ3} can be solved by dividing $B + k^2$ on both sides and integrating w.r.t. $k$ from $0$ to $\infty$:
\begin{equation}
    \int_0^\infty \frac{\hat{h}_{e,c}(k)}{B+k^2} \: \mathrm{d} k = \frac{c}{B+1} + \int_0^\infty k \hat{w}_c(k) \hat{g}_{e,c}(k) \: \mathrm{d} k - \alpha \int_0^\infty \frac{\hat{w}_c(k)}{B+k^2} \: \mathrm{d} k \cdot \int_0^\infty \frac{\hat{h}_{e,c}(k)}{B+k^2} \: \mathrm{d} k .
\end{equation}
Rearrange, we have:
\begin{subequations}
\begin{align}
\int_0^\infty \frac{\hat{h}_{e,c}(k)}{B+k^2} \: \mathrm{d} k &= \left( \frac{c}{B+1} + \int_0^\infty k \hat{w}_c(k) \hat{g}_{e,c}(k) \: \mathrm{d} k \right) \cdot \left( 1 + \alpha \int_0^\infty \frac{\hat{w}_c(k)}{B+k^2} \: \mathrm{d} k \right)^{-1} \nonumber \\
&= \left( 1 + \alpha I_3 \right)^{-1} \left( \frac{c}{B+1} + \int_0^\infty k \hat{w}_c(k) \hat{g}_{e,c}(k) \: \mathrm{d} k \right) ,
\end{align}
where
\begin{align}
I_3 &= \int_0^\infty \frac{\hat{w}_c(k)}{B+k^2} \: \mathrm{d} k \nonumber \\
&=
\frac{-\left((3 + B) \sqrt{3 + 4B} \upi \right) + 2 \sqrt{B} (3 + 2 B) \tan^{-1}\left(\sqrt{3 + 4 B}\right) + \sqrt{B (3 + 4 B)} \ln(1 + B)}{2 (1 + B) (3 + B) \sqrt{B (3 + 4 B)}} .
\end{align}
\end{subequations}
Thus,
\begin{align}\label{eq:hec}
\hat{h}_{e,c}(k) &= c \delta(k-1) + k (B+k^2) \hat{w}_c(k) \hat{g}_{e,c}(k) - \alpha \hat{w}_c(k) \int_0^\infty \frac{\hat{h}_{e,c}(k)}{B+k^2} \: \mathrm{d} k \nonumber \\
&= c \delta(k-1) + k (B+k^2) \hat{w}_c(k) \hat{g}_{e,c}(k) \nonumber \\
&\quad- \frac{\alpha}{1 + \alpha I_3} \cdot \hat{w}_c(k) \left( \frac{c}{B+1} + \int_0^\infty k \hat{w}_c(k) \hat{g}_{e,c}(k) \: \mathrm{d} k \right) , 
\end{align}
where $\hat{g}_{e,c}$ can be calculated by:
\begin{align}
    \hat{g}_{e,\pm,c} &= \mathcal{F}_c\left\{ g_{e, \pm} \right\} = \int_0^\infty g_{e,\pm}(x) \cos kx \: \mathrm{d} x , \nonumber \\
    &= \pm \mathrm{i}\left( \frac{1}{k} - \frac{1}{1 + k} \right) + \frac{ - (\sqrt{B} \pm \mathrm{i}) \mathrm{e}^{-2\mathrm{i}\zeta} + (-\sqrt{B} \pm \mathrm{i}) }{2 B} \left( \frac{1}{k} - \frac{k}{B + k^2} \right) .
\end{align}

Notice that on the positive $k$-axis, only the singularity terms in $h_c$ correspond to the terms in $h$ that propagate as $x \rightarrow \infty$. In this instance, there are two singularity terms on the positive $k$-axis in $h_{e,c}$: $(k-1)^{-1}$ and $\delta(k-1)$, which correspond to $\sin x$ and $\cos x$, respectively, at $x \rightarrow \infty$ in $h_e$. Since the equations for calculating $T$ and $R$ (eqs.\eqref{eq:T} and \eqref{eq:R}) only require the ratio $h_{e, -}/h_{e, +}$, it is sufficient to calculate this ratio using the coefficients of the $(k-1)^{-1}$ terms for $\hat{h}_{e,-,c}$ and $\hat{h}_{e,+,c}$. It can also be shown (albeit through another lengthy derivation) that the coefficient of $\delta(k-1)$ is equal to the coefficient of $(k-1)^{-1}$; we omit that proof for brevity. By substituting the expressions for $\alpha$, $\hat{w}_{c}$, and $\hat{g}_{e,c}$ into \eqref{eq:hec} and extracting the coefficients of the $(k-1)^{-1}$ terms, we obtain the following expression for $C_e$: 
\begin{equation}
    C_e = \lim_{x\rightarrow\infty}\frac{h_{e, -}}{h_{e, +}} = - \frac{X - \mathrm{i}}{X + \mathrm{i}} ,
\end{equation}
where
\begin{equation}\label{eq:app_X}
    X = \frac{1}{\upi} \left( \frac{3+2B}{\sqrt{3+4B}} \tan^{-1} \sqrt{3+4B} + \frac{1}{2} \ln (1+B) \right) + \mathrm{i}\frac{(3+B) \tan \zeta}{2\sqrt{B}} .
\end{equation}

Similarly, $h_o$ can be analyzed by applying a Fourier Sine Transform (FST) as follows
\begin{equation}
    \hat{f}_s(k) = \mathcal{F}_s\left\{ f \right\} = \int_0^\infty f(x) \sin kx \: \mathrm{d} x ,
\end{equation}
to \eqref{eq:heopmequ_appB}. The approach remains consistent, albeit the convolution theorem for FST differs:
\begin{equation}
    2 \mathcal{F}_s\left\{ f \right\} \mathcal{F}_c\left\{ g \right\} = \mathcal{F}_s\left\{ \int_0^\infty f(x') \left[ g(|x - x'|) - g(x + x') \right] \: \mathrm{d} x' \right\}.
\end{equation}
The details are omitted. The calculated expression for $C_o$:
\begin{equation}
     C_o = \lim_{x\rightarrow\infty} \frac{h_{o, -}}{h_{o, +}} = \frac{Y + \mathrm{i}}{Y - \mathrm{i}} ,
\end{equation}
where
\begin{equation}\label{eq:app_Y}
    Y = \frac{1}{\upi} \left( \frac{3+6B+2B^2}{\sqrt{3+4B}} \tan^{-1} \sqrt{3+4B} - \frac{1}{2} \ln (1+B) \right) - \mathrm{i}\frac{(3+B) \sqrt{B}}{2 \tan \zeta} .
\end{equation}

From \eqref{eq:T} and \eqref{eq:R}, we have the theoretical results for the transmission $T$ and the reflection $R$ coefficients of the infinitesimal barrier ($r \ll 1$) case:
\begin{subequations}
\begin{align}
    T &= -\frac{1}{2} (C_e - C_o) = \frac{1}{2} \left( \frac{X-\mathrm{i}}{X+\mathrm{i}} + \frac{Y+\mathrm{i}}{Y-\mathrm{i}} \right) , \label{eq:app_T} \\
    R &= -\frac{1}{2} (C_e + C_o) = \frac{1}{2} \left( \frac{X-\mathrm{i}}{X+\mathrm{i}} - \frac{Y+\mathrm{i}}{Y-\mathrm{i}} \right) . \label{eq:app_R}
\end{align}
\end{subequations}

These expressions are consistent with expressions from \citet{ref:Hocking-1987}, see Appendix~\ref{app:hocking} for details.

\section{Expressions for $T$, $R$ in Infinitesimal Barrier Case \citep{ref:Hocking-1987}}\label{app:hocking}

\citet{ref:Hocking-1987} considered scattering of capillary-gravity waves from an infinitesimally thin barrier of a finite immersed depth. In the limit of a zero depth, an analytical results followed from \citet{ref:Hocking-1987} for the transmission and reflection coefficients, $T$ and $R$. Here we show that these results agree with our analytical solutions of $T$ and $R$, \eqref{eq:T+R_infi}, derived for a cylindrical barrier of a zero cross section. 

The coefficients $T$ and $R$ by \citet{ref:Hocking-1987} were derived in terms of the inverse Bond number
\begin{equation}
    K = \frac{\sigma k^2}{\rho g},
    \label{eq:app_K}
\end{equation}
and a dimensionless slip coefficient
\begin{equation}
    \lambda = \sqrt{\frac{k}{g}} \Lambda'.
    \label{eq:app_lambda}
\end{equation}
In terms of $K$ and $\lambda$, the coefficients were given as
\begin{subequations}
\begin{align}
    T &= \frac{\mathrm{i}(\alpha - \beta)}{(1-\mathrm{i}\alpha) (1-\mathrm{i}\beta)}, \label{eq:HockingT} \\
    R &= \frac{1 + \alpha \beta}{(1-\mathrm{i}\alpha) (1-\mathrm{i}\beta)}, \label{eq:HockingR}
\end{align}
where
\begin{align}
    \alpha &= \frac{2 K}{(1+3K)(J_1(K)-\mathrm{i}\lambda(1+K)^{-1/2})}, \label{eq:app_alpha} \\
    \beta &= J_2(K)+\mathrm{i}(1+3K)(1+K)^{1/2}/2K\lambda, \label{eq:app_beta}
\end{align}
and
\begin{align}
    J_1(K) &= \frac{2K}{\upi(1+3K)} \left\{ \frac{1}{2}\ln\frac{K+1}{K} + \frac{3K+2}{K^{1/2}(3K+4)^{1/2}}\tan^{-1} \left( \frac{3K+4}{K} \right)^{1/2} \right\} , \\
    J_2(K) &= \frac{1}{\upi K} \left\{ -\frac{K}{2} \ln\frac{K+1}{K} + \frac{3K^2+6K+2}{K^{1/2}(3K+4)^{1/2}}\tan^{-1} \left( \frac{3K+4}{K} \right)^{1/2} \right\}, \label{eq:app_alpha_end}
\end{align}
\end{subequations}
in which a missing factor of $1/\pi$ in the original expression for $J_1(K)$ has been included, as identified by \citet{zhang2013}. 

Hocking's parameters $K$ and $\lambda$ relate to our Bond number $B$ and slip parameter $\zeta$ via
\begin{equation}
    K = \frac{1}{B}, \quad 
    \lambda  = \frac{\sqrt{1+B}}{B} \tan \zeta. 
    \label{eq:app_K_B_lambda_zeta}
\end{equation}
Substituting \eqref{eq:app_K_B_lambda_zeta} into \eqref{eq:app_alpha}--\eqref{eq:app_alpha_end}, the $\alpha$ and $\beta$ reduce to functions of the Bond number $B$ and $\zeta$ as:
\begin{align}
    \alpha &= \left\{ \frac{1}{\upi} \left( \frac{3+2B}{\sqrt{3+4B}} \tan^{-1} \sqrt{3+4B} + \frac{1}{2} \ln (1+B) \right) - \mathrm{i}\frac{(3+B) \tan \zeta}{2\sqrt{B}} \right\}^{-1} , \\
    \beta &= \frac{1}{\upi} \left( \frac{3+6B+2B^2}{\sqrt{3+4B}} \tan^{-1} \sqrt{3+4B} - \frac{1}{2} \ln (1+B) \right) + \mathrm{i}\frac{(3+B) \sqrt{B}}{2 \tan \zeta} .
\end{align}
Comparing them with our expressions of $X$ and $Y$ in \eqref{eq:T+R_infi}, we identify that
\begin{equation}
        \alpha = \frac{1}{X^*}, \quad \beta = Y^*, \label{eq:app_sub}
\end{equation}
where the stars denote complex conjugates. By substituting \eqref{eq:app_sub} into \eqref{eq:HockingT} and \eqref{eq:HockingR}, Hocking's $T$ and $R$ reduce to
\begin{subequations}
    \begin{align}
    T &
    = \frac{\mathrm{i}(1-X^*Y^*)}{(X^*-\mathrm{i})(1-\mathrm{i}Y^*)}
    = \left[\frac{XY-1}{(X+\mathrm{i})(Y-\mathrm{i})} \right]^*
    = \frac{1}{2} \left( \frac{X-\mathrm{i}}{X+\mathrm{i}} + \frac{Y+\mathrm{i}}{Y-\mathrm{i}} \right)^*, \\    
    R &
    = \frac{X^* + Y^*}{(X^*-\mathrm{i}) (1-\mathrm{i}Y^*)} 
    = \left[ \frac{-\mathrm{i}(X + Y)}{(X+\mathrm{i}) (Y-\mathrm{i})} \right]^* = \frac{1}{2} \left( \frac{X-\mathrm{i}}{X+\mathrm{i}} - \frac{Y+\mathrm{i}}{Y-\mathrm{i}} \right)^*.
\end{align}

These expressions for $T$ and $R$ are the complex conjugates of the expressions given in \eqref{eq:T+R_infi}. Since \citet{ref:Hocking-1987} uses the time convention $\exp(\mathrm{i}\omega t)$, while we use $\exp(-\mathrm{i}\omega t)$, Hocking's coefficients $T$ and $R$  are defined as the complex conjugates of our expressions. By using the complex conjugates, the two analyses agree with each other in this limiting case of an infinitesimal barrier.

\end{subequations}

\section{Table for Gravity Waves Scattering Comparison}\label{app:gravity}
See table attached.
\begin{table}
\centering
\resizebox{0.55\textwidth}{!}{
\begin{tabular}{|c|l|l|r|r|}
\hline
\multicolumn{1}{|c|}{$r$} & \multicolumn{1}{c|}{$|R|$} & \multicolumn{1}{c|}{$\arg R$} & \multicolumn{1}{c|}{$|T|$} & \multicolumn{1}{c|}{$\arg T$} \\
\hline
0.01 & \multicolumn{1}{l|}{$\begin{array}{l} 1.9650 (-2) \\ 1.9880 (-2) \end{array}$} & \multicolumn{1}{r|}{$\begin{array}{r} -1.5898 \\ -1.5900 \end{array}$} & \multicolumn{1}{l|}{$\begin{array}{l} 9.9980 (-1) \\ 9.9980 (-1) \end{array}$} & \multicolumn{1}{r|}{$\begin{array}{r} -0.0190 \\ -0.0193 \end{array}$} \\
0.02 & \multicolumn{1}{l|}{$\begin{array}{l} 3.8960 (-2) \\ 3.9260 (-2) \end{array}$} & \multicolumn{1}{r|}{$\begin{array}{r} -1.6073 \\ -1.6076 \end{array}$} & \multicolumn{1}{l|}{$\begin{array}{l} 9.9920 (-1) \\ 9.9920 (-1) \end{array}$} & \multicolumn{1}{r|}{$\begin{array}{r} -0.0365 \\ -0.0368 \end{array}$} \\
0.03 & \multicolumn{1}{l|}{$\begin{array}{l} 5.8160 (-2) \\ 5.8480 (-2) \end{array}$} & \multicolumn{1}{r|}{$\begin{array}{r} -1.6235 \\ -1.6238 \end{array}$} & \multicolumn{1}{l|}{$\begin{array}{l} 9.9830 (-1) \\ 9.9830 (-1) \end{array}$} & \multicolumn{1}{r|}{$\begin{array}{r} -0.0527 \\ -0.0530 \end{array}$} \\
0.04 & \multicolumn{1}{l|}{$\begin{array}{l} 7.7340 (-2) \\ 7.7650 (-2) \end{array}$} & \multicolumn{1}{r|}{$\begin{array}{r} -1.6385 \\ -1.6388 \end{array}$} & \multicolumn{1}{l|}{$\begin{array}{l} 9.9700 (-1) \\ 9.9700 (-1) \end{array}$} & \multicolumn{1}{r|}{$\begin{array}{r} -0.0677 \\ -0.0680 \end{array}$} \\
0.05 & \multicolumn{1}{l|}{$\begin{array}{l} 9.6560 (-2) \\ 9.6850 (-2) \end{array}$} & \multicolumn{1}{r|}{$\begin{array}{r} -1.6524 \\ -1.6527 \end{array}$} & \multicolumn{1}{l|}{$\begin{array}{l} 9.9530 (-1) \\ 9.9530 (-1) \end{array}$} & \multicolumn{1}{r|}{$\begin{array}{r} -0.0816 \\ -0.0819 \end{array}$} \\
0.06 & \multicolumn{1}{l|}{$\begin{array}{l} 1.1590 (-1) \\ 1.1610 (-1) \end{array}$} & \multicolumn{1}{r|}{$\begin{array}{r} -1.6653 \\ -1.6656 \end{array}$} & \multicolumn{1}{l|}{$\begin{array}{l} 9.9330 (-1) \\ 9.9320 (-1) \end{array}$} & \multicolumn{1}{r|}{$\begin{array}{r} -0.0945 \\ -0.0948 \end{array}$} \\
0.07 & \multicolumn{1}{l|}{$\begin{array}{l} 1.3530 (-1) \\ 1.3550 (-1) \end{array}$} & \multicolumn{1}{r|}{$\begin{array}{r} -1.6773 \\ -1.6775 \end{array}$} & \multicolumn{1}{l|}{$\begin{array}{l} 9.9080 (-1) \\ 9.9080 (-1) \end{array}$} & \multicolumn{1}{r|}{$\begin{array}{r} -0.1065 \\ -0.1067 \end{array}$} \\
0.08 & \multicolumn{1}{l|}{$\begin{array}{l} 1.5480 (-1) \\ 1.5490 (-1) \end{array}$} & \multicolumn{1}{r|}{$\begin{array}{r} -1.6884 \\ -1.6886 \end{array}$} & \multicolumn{1}{l|}{$\begin{array}{l} 9.8800 (-1) \\ 9.8790 (-1) \end{array}$} & \multicolumn{1}{r|}{$\begin{array}{r} -0.1176 \\ -0.1178 \end{array}$} \\
0.09 & \multicolumn{1}{l|}{$\begin{array}{l} 1.7440 (-1) \\ 1.7450 (-1) \end{array}$} & \multicolumn{1}{r|}{$\begin{array}{r} -1.6987 \\ -1.6988 \end{array}$} & \multicolumn{1}{l|}{$\begin{array}{l} 9.8470 (-1) \\ 9.8470 (-1) \end{array}$} & \multicolumn{1}{r|}{$\begin{array}{r} -0.1279 \\ -0.1280 \end{array}$} \\
0.1 & \multicolumn{1}{l|}{$\begin{array}{l} 1.9420 (-1) \\ 1.9420 (-1) \end{array}$} & \multicolumn{1}{r|}{$\begin{array}{r} -1.7081 \\ -1.7082 \end{array}$} & \multicolumn{1}{l|}{$\begin{array}{l} 9.8100 (-1) \\ 9.8100 (-1) \end{array}$} & \multicolumn{1}{r|}{$\begin{array}{r} -0.1373 \\ -0.1374 \end{array}$} \\
0.2 & \multicolumn{1}{l|}{$\begin{array}{l} 3.9580 (-1) \\ 3.9520 (-1) \end{array}$} & \multicolumn{1}{r|}{$\begin{array}{r} -1.7678 \\ -1.7677 \end{array}$} & \multicolumn{1}{l|}{$\begin{array}{l} 9.1830 (-1) \\ 9.1860 (-1) \end{array}$} & \multicolumn{1}{r|}{$\begin{array}{r} -0.1970 \\ -0.1969 \end{array}$} \\
0.3 & \multicolumn{1}{l|}{$\begin{array}{l} 5.8570 (-1) \\ 5.8460 (-1) \end{array}$} & \multicolumn{1}{r|}{$\begin{array}{r} -1.7906 \\ -1.7906 \end{array}$} & \multicolumn{1}{l|}{$\begin{array}{l} 8.1050 (-1) \\ 8.1130 (-1) \end{array}$} & \multicolumn{1}{r|}{$\begin{array}{r} -0.2198 \\ -0.2198 \end{array}$} \\
0.4 & \multicolumn{1}{l|}{$\begin{array}{l} 7.3690 (-1) \\ 7.3560 (-1) \end{array}$} & \multicolumn{1}{r|}{$\begin{array}{r} -1.8118 \\ -1.8120 \end{array}$} & \multicolumn{1}{l|}{$\begin{array}{l} 6.7600 (-1) \\ 6.7740 (-1) \end{array}$} & \multicolumn{1}{r|}{$\begin{array}{r} -0.2410 \\ -0.2412 \end{array}$} \\
0.5 & \multicolumn{1}{l|}{$\begin{array}{l} 8.4030 (-1) \\ 8.3910 (-1) \end{array}$} & \multicolumn{1}{r|}{$\begin{array}{r} -1.8550 \\ -1.8552 \end{array}$} & \multicolumn{1}{l|}{$\begin{array}{l} 5.4210 (-1) \\ 5.4400 (-1) \end{array}$} & \multicolumn{1}{r|}{$\begin{array}{r} -0.2842 \\ -0.2844 \end{array}$} \\
0.6 & \multicolumn{1}{l|}{$\begin{array}{l} 9.0460 (-1) \\ 9.0360 (-1) \end{array}$} & \multicolumn{1}{r|}{$\begin{array}{r} -1.9277 \\ -1.9279 \end{array}$} & \multicolumn{1}{l|}{$\begin{array}{l} 4.2630 (-1) \\ 4.2840 (-1) \end{array}$} & \multicolumn{1}{r|}{$\begin{array}{r} -0.3569 \\ -0.3571 \end{array}$} \\
0.7 & \multicolumn{1}{l|}{$\begin{array}{l} 9.4270 (-1) \\ 9.4190 (-1) \end{array}$} & \multicolumn{1}{r|}{$\begin{array}{r} -2.0275 \\ -2.0278 \end{array}$} & \multicolumn{1}{l|}{$\begin{array}{l} 3.3370 (-1) \\ 3.3590 (-1) \end{array}$} & \multicolumn{1}{r|}{$\begin{array}{r} -0.4568 \\ -0.4570 \end{array}$} \\
0.8 & \multicolumn{1}{l|}{$\begin{array}{l} 9.6500 (-1) \\ 9.6450 (-1) \end{array}$} & \multicolumn{1}{r|}{$\begin{array}{r} -2.1495 \\ -2.1495 \end{array}$} & \multicolumn{1}{l|}{$\begin{array}{l} 2.6210 (-1) \\ 2.6420 (-1) \end{array}$} & \multicolumn{1}{r|}{$\begin{array}{r} -0.5772 \\ -0.5787 \end{array}$} \\
0.9 & \multicolumn{1}{l|}{$\begin{array}{l} 9.7830 (-1) \\ 9.7780 (-1) \end{array}$} & \multicolumn{1}{r|}{$\begin{array}{r} -2.2878 \\ -2.2877 \end{array}$} & \multicolumn{1}{l|}{$\begin{array}{l} 2.0730 (-1) \\ 2.0930 (-1) \end{array}$} & \multicolumn{1}{r|}{$\begin{array}{r} -0.7170 \\ -0.7169 \end{array}$} \\
1.0 & \multicolumn{1}{l|}{$\begin{array}{l} 9.8620 (-1) \\ 9.8590 (-1) \end{array}$} & \multicolumn{1}{r|}{$\begin{array}{r} -2.4385 \\ -2.4383 \end{array}$} & \multicolumn{1}{l|}{$\begin{array}{l} 1.6550 (-1) \\ 1.6730 (-1) \end{array}$} & \multicolumn{1}{r|}{$\begin{array}{r} -0.8677 \\ -0.8675 \end{array}$} \\
2.0 & \multicolumn{1}{l|}{$\begin{array}{l} 9.9960 (-1) \\ 9.9960 (-1) \end{array}$} & \multicolumn{1}{r|}{$\begin{array}{r} 2.0608 \\ 2.0614 \end{array}$} & \multicolumn{1}{l|}{$\begin{array}{l} 2.6630 (-2) \\ 2.7420 (-2) \end{array}$} & \multicolumn{1}{r|}{$\begin{array}{r} -2.6516 \\ -2.6510 \end{array}$} \\
3.0 & \multicolumn{1}{l|}{$\begin{array}{l} 1.0000 \\ 1.0000 \end{array}$} & \multicolumn{1}{r|}{$\begin{array}{r} 0.1315 \\ 0.1321 \end{array}$} & \multicolumn{1}{l|}{$\begin{array}{l} 7.1320 (-3) \\ 7.6250 (-3) \end{array}$} & \multicolumn{1}{r|}{$\begin{array}{r} 1.7023 \\ 1.7031 \end{array}$} \\
4.0 & \multicolumn{1}{l|}{$\begin{array}{l} 1.0000 \\ 1.0000 \end{array}$} & \multicolumn{1}{r|}{$\begin{array}{r} -1.8325 \\ -1.8319 \end{array}$} & \multicolumn{1}{l|}{$\begin{array}{l} 2.5650 (-3) \\ 2.9560 (-3) \end{array}$} & \multicolumn{1}{r|}{$\begin{array}{r} 0.2617 \\ -0.2608 \end{array}$} \\
5.0 & \multicolumn{1}{l|}{$\begin{array}{l} 1.0000 \\ 1.0000 \end{array}$} & \multicolumn{1}{r|}{$\begin{array}{r} 2.4728 \\ 2.4733 \end{array}$} & \multicolumn{1}{l|}{$\begin{array}{l} 1.1170 (-3) \\ 1.4610 (-3) \end{array}$} & \multicolumn{1}{r|}{$\begin{array}{r} -2.2396 \\ -2.2386 \end{array}$} \\
6.0 & \multicolumn{1}{l|}{$\begin{array}{l} 1.0000 \\ 1.0000 \end{array}$} & \multicolumn{1}{r|}{$\begin{array}{r} 0.4877 \\ 0.4881 \end{array}$} & \multicolumn{1}{l|}{$\begin{array}{l} 5.5630 (-4) \\ 8.7070 (-4) \end{array}$} & \multicolumn{1}{r|}{$\begin{array}{r} 2.0585 \\ 2.0596 \end{array}$} \\
7.0 & \multicolumn{1}{l|}{$\begin{array}{l} 1.0000 \\ 1.0000 \end{array}$} & \multicolumn{1}{r|}{$\begin{array}{r} -1.5016 \\ -1.5011 \end{array}$} & \multicolumn{1}{l|}{$\begin{array}{l} 3.0530 (-4) \\ 5.9760 (-4) \end{array}$} & \multicolumn{1}{r|}{$\begin{array}{r} 0.0693 \\ 0.0705 \end{array}$} \\
8.0 & \multicolumn{1}{l|}{$\begin{array}{l} 1.0000 \\ 1.0000 \end{array}$} & \multicolumn{1}{r|}{$\begin{array}{r} 2.7898 \\ 2.7902 \end{array}$} & \multicolumn{1}{l|}{$\begin{array}{l} 1.8050 (-4) \\ 4.5430 (-4) \end{array}$} & \multicolumn{1}{r|}{$\begin{array}{r} -1.9226 \\ -1.9213 \end{array}$} \\
9.0 & \multicolumn{1}{l|}{$\begin{array}{l} 1.0000 \\ 1.0000 \end{array}$} & \multicolumn{1}{r|}{$\begin{array}{r} 0.7962 \\ 0.7966 \end{array}$} & \multicolumn{1}{l|}{$\begin{array}{l} 1.1310 (-4) \\ 3.7100 (-4) \end{array}$} & \multicolumn{1}{r|}{$\begin{array}{r} 2.3670 \\ 2.3683 \end{array}$} \\
10.0 & \multicolumn{1}{l|}{$\begin{array}{l} 1.0000 \\ 1.0000 \end{array}$} & \multicolumn{1}{r|}{$\begin{array}{r} -1.1986 \\ -1.1982 \end{array}$} & \multicolumn{1}{l|}{$\begin{array}{l} 7.4280 (-5) \\ 3.1810 (-4) \end{array}$} & \multicolumn{1}{r|}{$\begin{array}{r} 0.3722 \\ 0.3734 \end{array}$} \\
\hline
\end{tabular}}
\caption{Comparison of our calculated results and results from \citet{martin1983scattering} for gravity wave scattering, with martin \& Dixon's at the top of each cell and ours at the bottom.}\label{tb:grav-compare}
\end{table}

\end{document}